\documentclass[usenatbib,useAMS]{mnras}

\usepackage{array}

\usepackage{enumerate}

\usepackage{aas_macros}

\usepackage{amsmath}
\usepackage{autobreak}
\usepackage{mathtools}
\usepackage{float}

\usepackage{upgreek}

\usepackage{xcolor}

\usepackage{txfonts}

\usepackage[applemac]{inputenc}
\usepackage[T1]{fontenc} 

\usepackage[english]{babel}
\usepackage{booktabs}
\usepackage{multirow}

\usepackage{amssymb,amsmath}
\usepackage{pifont}

\usepackage{amstext}
\usepackage{graphicx}
\usepackage[tableposition=top]{caption}
\usepackage{subcaption}
\usepackage[np,noautolanguage]{numprint}

\usepackage{scalefnt}
\usepackage{color}

\usepackage{mathrsfs}

\hypersetup{breaklinks, colorlinks, linktocpage=true, linkcolor=blue, citecolor=blue, filecolor=black, urlcolor=blue}

\usepackage{relsize}

\usepackage{etoolbox}
\makeatletter

\patchcmd{\NAT@citex}
  {\@citea\NAT@hyper@{%
     \NAT@nmfmt{\NAT@nm}%
     \hyper@natlinkbreak{\NAT@aysep\NAT@spacechar}{\@citeb\@extra@b@citeb}%
     \NAT@date}}
  {\@citea\NAT@nmfmt{\NAT@nm}%
   \NAT@aysep\NAT@spacechar\NAT@hyper@{\NAT@date}}{}{}

\patchcmd{\NAT@citex}
  {\@citea\NAT@hyper@{%
     \NAT@nmfmt{\NAT@nm}%
     \hyper@natlinkbreak{\NAT@spacechar\NAT@@open\if*#1*\else#1\NAT@spacechar\fi}%
       {\@citeb\@extra@b@citeb}%
     \NAT@date}}
  {\@citea\NAT@nmfmt{\NAT@nm}%
   \NAT@spacechar\NAT@@open\if*#1*\else#1\NAT@spacechar\fi\NAT@hyper@{\NAT@date}}
  {}{}

\makeatother

\bibpunct{(}{)}{;}{a}{}{,} 

\usepackage{threeparttable}

\numberwithin{equation}{section}

		\usepackage{stfloats}

\newcolumntype{L}[1]{>{\raggedright\let\newline\\\arraybackslash\hspace{0pt}}m{#1}}
\newcolumntype{C}[1]{>{\centering\let\newline\\\arraybackslash\hspace{0pt}}m{#1}}
\newcolumntype{R}[1]{>{\raggedleft\let\newline\\\arraybackslash\hspace{0pt}}m{#1}}

\renewcommand{\bmath}[1]{\mbox{ \boldmath $\!#1\!$ \unboldmath}}

\newcommand{\overbar}[1]{\mkern 1.5mu\overline{\mkern-1.5mu#1\mkern-1.5mu}\mkern 1.5mu}

\newcommand{\txn}[1]{\textnormal{#1}}

\newcommand{\conditional}[2]{\hbox{$\txn{P}(#1 \mid #2)$}}
\newcommand{\range}[3]{\hbox{$#1 \sim #2 \,\txn{--}\, #3$}}
\newcommand{\IRACone}{\hbox{$\txn{IR}_1$}}
\newcommand{\IRACtwo}{\hbox{$\txn{IR}_2$}}
\newcommand{\IRACcolour}{\hbox{$\IRACone - \IRACtwo$}}

\newcommand{\Planck}{\textit{Planck}}

\LetLtxMacro{\oldtextsc}{\textsc}
\renewcommand{\textsc}[1]{\oldtextsc{\scalefont{1.2}#1}}

\newcommand{\beagle}{\textsc{beagle}}
\newcommand{\multinest}{\textsc{multinest}}

\renewcommand{\lambda}{\uplambda}

\newcommand{\sqarcmin}{\hbox{$\txn{arcmin}^2$}}
\newcommand{\invsqarcmin}{\hbox{$\txn{arcmin}^{-2}$}}

\newcommand{\upsigmao}{\hbox{$\upsigma_\txn{obs, i}$}}

\newcommand{\Hubble}{{\it Hubble}}

\newcommand{\Spitzer}{{\it Spitzer}}
\newcommand{\JWST}{{\it JWST}}
\newcommand{\HST}{{\it HST}}

\newcommand{\npt}{\hbox{$\txn{n}_\txn{p}$}}
\newcommand{\Nobj}{\hbox{$\txn{N}_\txn{obj}$}}
\newcommand{\Sobj}{\hbox{$\upsigma_\txn{obj}$}}

\newcommand{\R}{\hbox{$\txn{R}$}}
\newcommand{\SN}{\hbox{$\txn{S}/\txn{N}$}}
\newcommand{\EW}{\hbox{$\txn{EW}$}}
\newcommand{\Gyr}{\hbox{$\txn{Gyr}$}}

\newcommand{\Nlines}{\hbox{$\txn{N}$}}

\newcommand{\yrInv}{\hbox{$\txn{yr}^{-1}$}}

\newcommand{\thetab}{\hbox{$\bmath{\Theta}$}}
\newcommand{\Db}{\hbox{$\bmath{D}$}}

\newcommand{\M}{\hbox{$\txn{M}$}}

\newcommand{\zspec}{\hbox{$z_\txn{spec}$}}

\newcommand{\Mstar}{\hbox{$\M_{\ast}$}}
\newcommand{\Msun}{\hbox{$\M_{\sun}$}}

\newcommand{\logM}{\hbox{$\log(\M/\Msun)$}}

\newcommand{\Muv}{\hbox{$M_\textsc{uv}$}}

\newcommand{\MsunyrInv}{\hbox{$\Msun\,\yrInv$}}

\newcommand{\Omegam}{\hbox{$\Omega_{\mathrm{m}}$}}

\newcommand{\Omegal}{\hbox{$\Omega_{\Lambda}$}}

\renewcommand{\t}{\hbox{$t$}}
\newcommand{\tprime}{\hbox{$\t^{\prime}$}}

\newcommand{\tformm}{\hbox{$\t_{\scriptscriptstyle \txn{form}}^{\scriptscriptstyle \txn{max}}$}}
\newcommand{\zformm}{\hbox{$z_{\scriptscriptstyle \txn{form}}^{\scriptscriptstyle \txn{max}}$}}

\newcommand{\logOH}{\hbox{$12 + \log (\txn{O}/\txn{H})$}}
\newcommand{\Z}{\hbox{$\txn{Z}$}}
\newcommand{\Zgas}{\hbox{$\Z_\txn{gas}$}}
\newcommand{\Zsun}{\hbox{$\Z_\odot$}}
\newcommand{\Zyoung}{\hbox{$\Z_\txn{young}$}}

\newcommand{\tausfr}{\hbox{$\uptau_\textsc{sfr}$}}

\newcommand{\Tuniverse}{\hbox{$\txn{t}_{\textsc{u},z}$}}

\newcommand{\sfrc}{\hbox{$\uppsi$}}
\newcommand{\ssfr}{\hbox{${\uppsi_\textsc{s}}$}}

\newcommand{\nObjBin}[1]{\newline ({#1})}

\newcommand{\re}{\hbox{$r_\txn{e}$}}

\newcommand{\Us}{\hbox{$U_\txn{S}$}}
\newcommand{\logUs}{\hbox{$\log \Us$}}
\newcommand{\xid}{\hbox{$\upxi_\txn{d}$}}
\newcommand{\Zism}{\hbox{$\Z_\textsc{ism}$}}

\newcommand{\lam}[1]{\hbox{$\uplambda#1$}}
\newcommand{\lamlam}[2]{\hbox{$\uplambda\uplambda#1,\!#2$}}

\newcommand{\magH}{\hbox{$m_\txn{F160W}$}}

\newcommand{\Hg}{\hbox{H$\upgamma$}}
\newcommand{\Ha}{\hbox{H$\upalpha$}}
\newcommand{\HeII}{\hbox{He\,{\sc ii}\lam{1640}}}
\newcommand{\OIIIuv}{\mbox{O\,{\sc iii]}\lamlam{1660}{1666}}}
\newcommand{\SiIII}{\mbox{Si\,{\sc iii]}\lamlam{1883}{1892}}}
\newcommand{\Hb}{\hbox{H$\beta$}}
\newcommand{\OII}{\mbox{[O\,{\sc ii]}\lamlam{3726}{3729}}}
\newcommand{\OIInoL}{\mbox{[O\,{\sc ii]}}}
\newcommand{\OIIa}{\mbox{[O\,{\sc ii]}\lam{3726}}}
\newcommand{\OIIb}{\mbox{[O\,{\sc ii]}\lam{3729}}}
\newcommand{\OIIInoL}{\mbox{[O\,{\sc iii]}}}
\newcommand{\OIII}{\mbox{[O\,{\sc iii]}\lamlam{4959}{5007}}}
\newcommand{\OIIIa}{\mbox{[O\,{\sc iii]}\lam{4959}}}
\newcommand{\OIIIb}{\mbox{[O\,{\sc iii]}\lam{5007}}}

\newcommand{\NII}{\mbox{[N\,{\sc ii]}\lamlam{6548}{6584}}}
\newcommand{\SII}{\mbox{[S\,{\sc ii]}\lamlam{6716}{6731}}}
\newcommand{\Lya}{\hbox{Ly$\upalpha$}}
\newcommand{\CIII}{\hbox{[C\,{\sc iii}]\lam1907+C\,{\sc iii}]\lam1909}}
\newcommand{\CIIIa}{\hbox{[C\,{\sc iii}]\lam1907}}
\newcommand{\CIIIb}{\hbox{C\,{\sc iii}]\lam1909}}

\newcommand{\HaNIISII}{\mbox{$\Ha+\txn{[N\,{\sc ii]}}+\txn{[S\,{\sc ii]}}$}}
\newcommand{\HbOIII}{\mbox{$\Hb+\txn{[O\,{\sc iii]}}$}}

\newcommand{\SNHb}{\mbox{ $\SN_{\scriptsize\Hb}$}}

\newcommand{\EWCIII}{\mbox{$\txn{W}_{\txn{C}\,\textsc{iii]}}$}}

\newcommand{\Hii}{\mbox{H\,{\sc ii}}}

\newcommand{\tauV}{\hbox{$\hat{\uptau}_{\scriptscriptstyle V}$}}
\newcommand{\mud}{\hbox{$\upmu$}}

\newcommand{\affilSpace}{-5pt}

\title[Simulating deep NIRSpec observations in the HUDF]{Simulating and interpreting deep observations in the Hubble Ultra Deep Field with the \JWST/NIRSpec low-resolution `prism'}
\author[J.~Chevallard]
{Jacopo~Chevallard$^{1}$\thanks{E-mail: chevallard@iap.fr}\thanks{ESA Research Fellow},
Emma~Curtis-Lake$^{2}$,
St\'ephane~Charlot$^{2}$,
Pierre~Ferruit$^{1}$,
\vspace{-8pt} \newauthor \vspace{-8pt}
Giovanna~Giardino$^{1}$,
Marijn~Franx$^{3}$,
Michael~V.~Maseda$^{3}$,
Ricardo Amorin$^{4,5}$, 
\vspace{-8pt} \newauthor \vspace{-8pt}
Santiago Arribas$^{6}$, 
Andy Bunker$^{7}$,
Stefano Carniani$^{5,6}$, 
Bernd Husemann$^{8}$, 
\vspace{-8pt} \newauthor \vspace{-8pt}
Peter Jakobsen$^{9}$, 
Roberto Maiolino$^{5,6}$, 
Janine Pforr$^{1}$, 
Timothy~D.~Rawle$^{10}$, 
\vspace{-8pt} \newauthor \vspace{-8pt} 
Hans-Walter Rix$^{8}$,
Renske Smit$^{5,6}$, 
Chris J. Willott$^{11}$
\\
\\
\vspace{\affilSpace} $^{1}$Scientific Support Office, Directorate of Science and Robotic Exploration, ESA/ESTEC, Keplerlaan 1, 2201 AZ Noordwijk, The Netherlands\\
\vspace{\affilSpace} $^{2}$Sorbonne Universit\'es, UPMC-CNRS, UMR7095, Institut d'Astrophysique de Paris, F-75014, Paris, France\\
\vspace{\affilSpace} $^{3}$Leiden Observatory, Leiden University, NL-2300 RA Leiden, Netherlands\\
\vspace{\affilSpace} $^{4}$Cavendish Laboratory, University of Cambridge, 19 J. J. Thomson Ave., Cambridge CB3 0HE, UK \\
\vspace{\affilSpace} $^{5}$Kavli Institute for Cosmology, University of Cambridge, Madingley Road, Cambridge CB3 0HA, UK \\
\vspace{\affilSpace} $^{6}$Departamento de Astrof\'isica, Centro de Astrobiolog\'ia, CSIC-INTA, Cra. de Ajalvir, 28850-Madrid, Spain \\
\vspace{\affilSpace} $^{7}$Department of Physics, University of Oxford, Oxford, UK \\
\vspace{\affilSpace} $^{8}$Max Planck Institute for Astronomy, K\"onigstuhl 17, D-69117 Heidelberg, Germany \\
\vspace{\affilSpace} $^{9}$Dark Cosmology Centre, Niels Bohr Institute, University of Copenhagen, Juliane Maries vej 30, DK-2100 Copenhagen, Denmark\\
\vspace{\affilSpace} $^{10}$European Space Agency, c/o STScI, 3700 San Martin Drive, Baltimore, MD 21218, USA \\
\vspace{\affilSpace} $^{11}$NRC Herzberg, 5071 West Saanich Rd, Victoria, BC V9E 2E7, Canada 
}

\begin{document}

\date{Submitted to MNRAS on }

\maketitle

\label{firstpage}

\begin{abstract}

The \textit{James Webb Space Telescope} (\JWST) will enable the detection of optical emission lines in galaxies spanning a broad range of luminosities out to redshifts $z\gtrsim10$. Measurements of key galaxy properties, such as star formation rate and metallicity, through these observations will provide unique insight into, e.g., the role of feedback from stars and active galactic nuclei (AGNs) in regulating galaxy evolution, the co-evolution of AGNs and host galaxies, the physical origin of the `main sequence' of star-forming galaxies and the contribution by star-forming galaxies to cosmic reionization. We present an original framework to simulate and analyse observations performed with the Near Infrared Spectrograph (NIRSpec) on board \JWST. We use the \beagle\ tool (BayEsian Analysis of GaLaxy sEds) to build a semi-empirical catalogue of galaxy spectra based on photometric spectral energy distributions (SEDs) of dropout galaxies in the \Hubble\ Ultra Deep Field (HUDF). We demonstrate that the resulting catalogue of galaxy spectra satisfies different types of observational constraints on high redshift galaxies, and use it as input to simulate NIRSpec/prism ($R\sim 100$) observations. We show that a single `deep' ($\sim100$ ks) NIRSpec/prism pointing in the HUDF will enable $\SN>3$ detections of multiple optical emission lines in $\sim30$ ($\sim60$) galaxies at $z\gtrsim6$ (\range{z}{4}{6}) down to $\magH \lesssim 30$ AB mag. Such observations will allow measurements of galaxy star formation rates, ionization parameters and gas-phase metallicities within factors of  1.5, mass-to-light ratios within a factor of 2, galaxy ages within a factor of 3 and $V$-band attenuation optical depths with a precision of 0.3.


\end{abstract}



\section{Introduction}\label{sec:intro}

Our picture of the formation and evolution of galaxies across cosmic time has improved significantly in the past 15 years. Galaxy surveys have provided researchers with a wealth of spectro-photometric data, at both low (e.g. SDSS, \citealt{York2000}) and high (e.g., GOODS, \citealt{Stanway2003}; GLARE, \citealt{Stanway2004,Stanway2007}; COSMOS, \citealt{Scoville2007}; HUDF, \citealt{Bunker2004, Beckwith2006, Bouwens2010, Bunker2010, McLure2010, Ellis2013, Illingworth2013}; CANDELS, \citealt{Grogin2011, Koekemoer2011}; VUDS, \citealt{LeFevre2015}; VANDELS, \citealt{McLure2017}) redshifts. On the theoretical front, computer simulations of galaxy formation (e.g. EAGLE, \citealt{Schaye2015}; IllustrisTNG, \citealt{Pillepich2017}) can now accurately predict several key galaxy properties, such as the redshift evolution of the galaxy stellar mass function and star formation rate density \citep[e.g.][]{Genel2014, Furlong2015}. Yet, many details of the physical processes driving the evolution of galaxies remain unknown. AGN feedback is a key ingredient of galaxy formation models (see \citealt{Somerville2015} and references therein), but observational evidence for AGN-driven `quenching' of star formation is ambiguous \citep[e.g.][]{Carniani2016, Suh2017}. Similarly, although several correlations exist between the physical properties of AGNs and their host galaxies (see \citealt{Kormendy2013} and references therein), a causal connection among these properties indicating a `co-evolution' of galaxies and AGNs has not been clearly demonstrated. The apparent existence of a tight relation between galaxy masses and star formation rates (i.e., `main sequence', \citealt{Brinchmann2004, Noeske2007}), and perhaps of a more `fundamental' relation involving gas-phase metallicity as well \citep{Mannucci2010}, can be an indication of self-regulation of star formation within galaxies \citep[e.g.][]{Lilly2013}, but the significance and redshift evolution of these relations remains unclear \citep[e.g.][]{Yabe2015, Telford2016}. The increasing relative abundance of UV-faint galaxies at redshift $z \gtrsim 6$ \citep[e.g.][]{Bouwens2015, Finkelstein2015} would support a picture in which low-mass star-forming galaxies provide the bulk of H-ionizing (LyC) photons required for cosmic reionization \citep[e.g.][]{Wilkins2011, Finkelstein2012, Robertson2013, Bouwens2015b}, but this conclusion relies on several assumptions about the ionizing emissivity of galaxies and escape fraction of LyC photons from galaxies.

Advancing our understanding of the above processes, and of many others, that drive the evolution of galaxies from the reionization epoch to the present day, requires measuring physical properties -- such as stellar masses, star formation rates, stellar and gas metallicities, stellar ages, gas ionization state, the dynamics of gas and stars -- for large  samples of galaxies, across a broad range of redshifts and galaxy stellar masses. While multi-band photometric campaigns have collected high quality SEDs for thousands of galaxies spanning a wide range of masses out to $z\gtrsim 8$ (e.g. HUDF, CANDELS), the availability of spectroscopic observations at $z\gtrsim 1$ is much more limited. At those redshifts, the strongest optical emission lines (e.g. \OII, \Hb, \OIII, \Ha) are shifted to near infrared wavelengths, where observations with ground-based telescopes are challenging because of regions of low atmospheric transmittance, bright sky background and contamination from bright, variable sky lines. Optical emission lines have been measured out to $z\sim3$ for large samples of galaxies using low-resolution slitless spectroscopy with \HST\ (e.g. WISP, \citealt{Atek2010}; 3D-HST, \citealt{Brammer2012}; GLASS, \citealt{Treu2015}; FIGS, \citealt{Pirzkal2017}). Higher-resolution spectroscopic surveys from the ground (e.g. VUDS, VANDELS) have mainly relied on multi-object spectrographs operating at optical wavelengths (e.g. VIMOS at VLT, \citealt{LeFevre2003}), hence are limited, at high redshift, to the measurement of rest-frame UV emission lines. These lines are intrinsically weaker than the optical ones (e.g., see table~5 of \citealt{Steidel2016}), and their interpretation is complicated by radiative transfer effects, dust attenuation and potential contamination from outflows.\footnote{The \Lya\ line is theoretically the most luminous emission line at UV and optical wavelengths, but being a resonant line it suffers from radiative transfer effects which affect its visibility and make its physical interpretation extremely challenging.}
Multi-object, ground-based spectrographs operating at near infrared wavelengths (e.g. MOSFIRE at Keck, \citealt{McLean2008}; KMOS at VLT, \citealt{Sharples2013}) permit the extension of rest-frame optical-emission-line measurements out to $z\sim4$, albeit only for galaxies with relatively bright emission lines (e.g. MOSDEF, \citealt{Kriek2015}; KBSS, \citealt{Steidel2014}). These measurements enabled, for example, new constraints on the dust attenuation properties of galaxies at $z\sim2$ \citep[e.g.][]{Reddy2015} and on the conditions of ionized gas (metal abundances, ionization state)  at \range{z}{2}{4} \citep[e.g.][]{Strom2017, Shapley2017}.

In the near future, the limitations of existing observatories to study the high redshift Universe will largely be overcome by the launch of the \textit{James Webb Space Telescope} (\JWST, \citealt{Gardner2006}). \JWST\ is expected to revolutionise our view of galaxies at $z\gtrsim 4$, and of low-mass galaxies at $1 \le z \le 4$, by providing a wide range of imaging and spectroscopic capabilities in the wavelength range \range{\lambda}{0.6}{28} \micron. In particular, the Near Infrared Spectrograph (NIRSpec, \citealt{Bagnasco2007, Birkmann2016}) onboard \JWST\ will increase by more than an order of magnitude the emission line sensitivity in the wavelength region \range{\lambda}{1}{2.5} \micron\ covered by existing ground-based instruments, while reaching an even greater sensitivity at wavelengths $\lambda \gtrsim 2.5$ \micron\ inaccessible with existing observatories. The multi-object spectroscopic capabilities of NIRSpec will enable the simultaneous measurement of up to $\sim 200$ galaxy spectra, allowing the observation of standard optical emission lines for large samples of galaxies out to $z\gtrsim 10$ (see Section~\ref{sec:external_constraints}). Combined with stellar mass measurements based on \JWST/NIRCam imaging (Near Infrared Camera, \citealt{Horner2004}), this will provide researchers with unique data to constrain the physical processes driving the evolution of galaxies at $z\gtrsim4$, the formation of the first galaxies at $z\gtrsim 10$, the way cosmic reionization proceeded in space and time and the contribution of different sources to the cosmic reionization budget. 

The uniqueness of \JWST/NIRSpec, however, also poses new challenges for the planning and interpretation of observations of high redshift galaxies. Unlike ground-based multi-object spectrographs, which typically require the user to define a `mask' of apertures on the sky, NIRSpec is equipped with a micro-shutter array (MSA, \citealt{Kutyrev2008}) providing a fixed grid of apertures on the sky. Optimising the NIRSpec MSA usage for a given scientific goal therefore requires a careful prioritization of the targets, as well as calculations to study spectral overlaps and truncations. The different filters and dispersers available on NIRSpec, covering the wavelength range \range{\lambda}{0.6}{5.3} \micron\ and spectral resolutions $R\sim100$, $\sim1000$ and $\sim 2700$, provide complementary information, but the optimal choice of filters, dispersers and exposure times depends on the scientific case, target properties and redshift range. Given the paucity of high quality galaxy spectra at $z\gtrsim 4$, NIRSpec data will open a largely unexplored space of observables, for both emission lines and stellar continuum studies. Maximising the information extracted from such data therefore requires models and approaches adapted to describing galaxies with a wide range of stellar populations and interstellar medium properties, likely extending well beyond those measured with existing observatories. Having adequate models and simulations is therefore critical, especially for the planning of Cycle 1 and Cycle 2 observations, as these will be largely based on spectroscopic follow-up of existing, pre-\JWST\ imaging data. 

In this paper, we build a physically-motivated, semi-empirical framework to simulate integrated galaxy spectra as could be observed at low spectral resolution with \JWST/NIRSpec in the \Hubble\ Ultra Deep Field (HUDF). We then use these simulated spectra to study our ability to constrain different galaxy physical parameters, such as star formation rates, ages, mass-to-light ratios, metallicities and properties of dust attenuation and ionised gas, for objects at redshift \range{z}{4}{8}. We adopt a self-consistent physical model which accounts for stellar and ionised gas emission, integrated in the state-of-the-art \beagle\ tool \citep{Chevallard2016}, to generate the input mock catalog of galaxy spectra and to fit the simulated observed spectra to retrieve the galaxy properties.

In Section~\ref{sec:semi_empirical}, we describe the catalogue of $z\gtrsim3$ dropouts of \citet{Bouwens2015} and our approach to fit the UV-to-near infrared photometry of these sources with the \beagle\ tool. We also present our method for associating model spectra with each dropout galaxy, and demonstrate how such a method produces spectra consistent with several, independent observables. In Section~\ref{sec:nirspec_simulations} we present our simulations of \JWST/NIRSpec observations and the full-spectrum fitting of these simulations with \beagle. In Section~\ref{sec:constraints} we examine the constraints on different galaxy physical parameters obtained by fitting the simulated NIRSpec data. In Section~\ref{sec:discussion} we discuss our results in the light of NIRSpec observational programs, in particular of the NIRSpec Guaranteed Time Observations (GTO) program, and future extensions of this work to cover a broader range of NIRSpec observing modes. Finally, we summarize our conclusions in Section~\ref{sec:conclusions}.

Throughout the paper, we express magnitudes in the AB system, adopt a zero-age solar metallicity $\Zsun = 0.017$ (corresponding to a present-day metallicity of 0.01524, see Table~3 of \citealt{Bressan2012}) and the latest constraints on cosmological parameters from  \Planck, i.e. $\Omegal = 0.6911$,  $\Omegam = 0.3089$ and $H_0 = 67.74$ \citep[see last column `TT,TE,EE+lowP+lensing+ext' of Table 4 of][]{Planck2015}. All the emission line equivalent widths in the text refer to rest-frame values.

\section{A semi-empirical catalogue of galaxy spectra in the HUDF}\label{sec:semi_empirical}

In this section we detail our approach to create a semi-empirical catalogue of high-redshift galaxy spectra. This catalogue is constructed by matching the predictions of a spectral evolution model to a large sample of $z\gtrsim3$ galaxies with multi-band \HST\ photometry, and represents the first of several steps involved in the process of creating and interpreting simulated NIRSpec observations. 
The adoption of a semi-empirical approach is motivated by the need to minimize the model dependence of our analysis, decoupling it from a particular theoretical approach (e.g. hydro-dynamic simulation \emph{vs} semi-analytic model) and its specific implementation. Also, anchoring our simulations to existing \HST\ photometry allows us to create and study NIRSpec observations which more closely resemble those that will be obtained early on in the \JWST\ mission, i.e. based on \HST-selected targets. Adopting a purely empirical approach based on existing spectroscopic observations, on the other hand, is not viable because of the widely different wavelength coverage and sensitivity of current observatories compared to \JWST/NIRSpec. Existing spectroscopic surveys targeting high redshift galaxies are often limited to observed wavelengths $\lambda \lesssim 1$ \micron\ (e.g. VVDS, \citealt{LeFevre2013}, VUDS, VANDELS), while surveys extending to $\lambda\sim2.5$ \micron\ (e.g. MOSDEF) target relatively bright ($\magH \le 25$) galaxies.
We note that the semi-empirical approach adopted here, which relies on observations of UV-selected galaxies, can miss populations of galaxies with little UV emission, such as very dusty star-forming galaxies and passively evolving ones. This UV selection reflects what will actually be used to prepare early \JWST/NIRSpec observations, before NIRCam observations of depth comparable to that of existing \HST\ observations become available. In future work, we will use the recently published mock galaxy catalogue of \citet{Williams2018} to analyse NIRSpec targets selected from deep NIRCam observations.

\subsection{Multi-band \HST\ photometry of dropout galaxies in the HUDF}\label{sec:XDF}

\begin{figure}
	\centering
	\resizebox{\hsize}{!}{\includegraphics{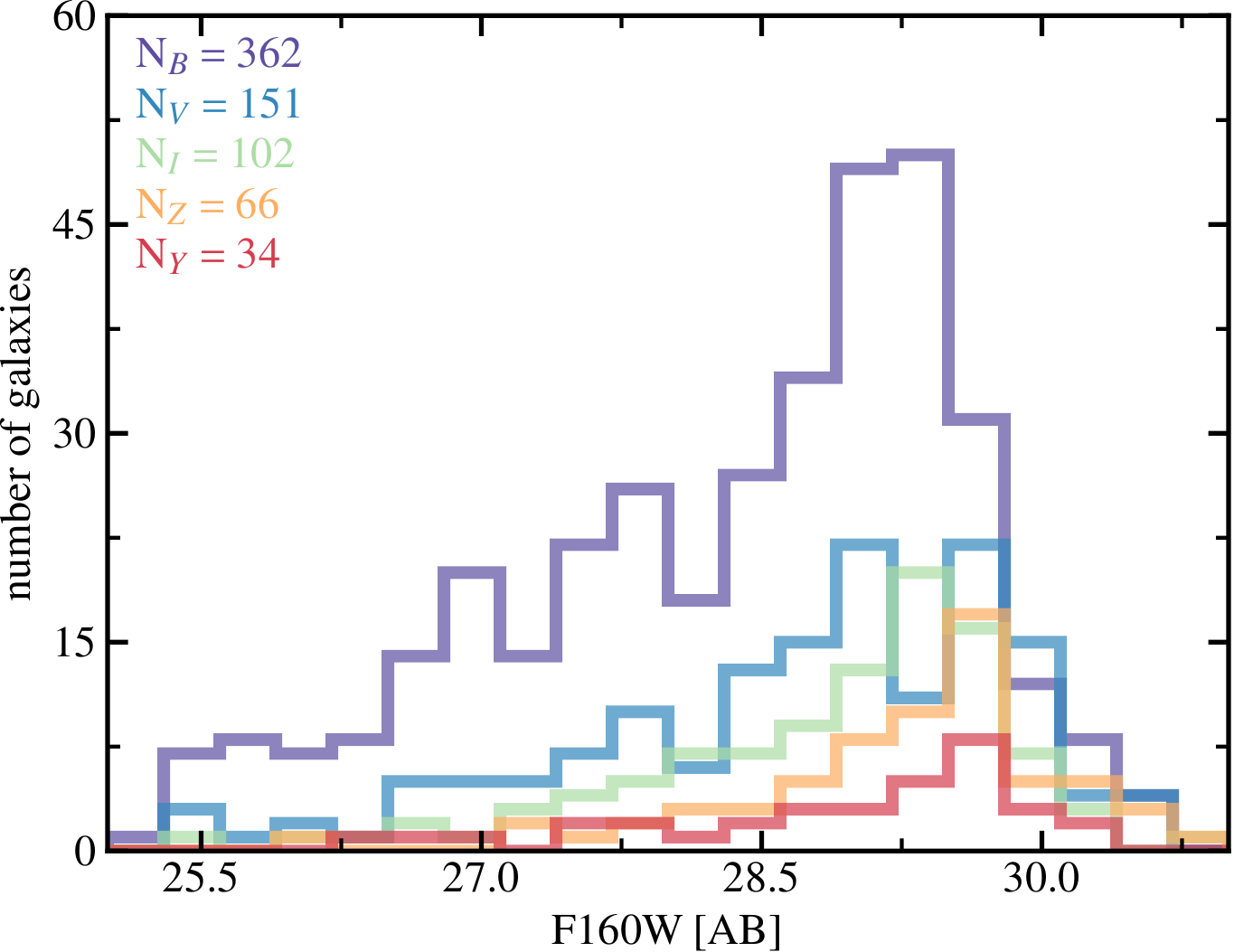}}
	\caption{Distributions of the WFC3/F160W magnitude for the $B$ ($z\sim 4$, purple line),  $V$ ($z\sim 5$, cyan),  $I$ ($z\sim 6$, green),  $Z$ ($z\sim 7$ objects, orange) and  $Y$ ($z\sim 8$ objects, red) dropout galaxies selected by \citet{Bouwens2015} in the HUDF.}
	\label{fig:XDF_mag}
\end{figure} 

In this work, we focus on galaxies at redshift $z\gtrsim 3$, for which \JWST/NIRSpec will enable measurements of standard optical emission lines (e.g. \Hb, \OIII, \Ha, \NII, \SII), which are largely inaccessible with existing ground-based telescopes. As shown by the initial works of \citet{Steidel1996} and \citet{Madau1996}, high redshift star-forming galaxies can be effectively selected from broad-band photometric data by using the Lyman break `dropout' technique. This  technique exploits the (nearly) complete absorption by neutral hydrogen of any light emitted by a galaxy blue-ward the Lyman limit (912 \AA). As the Lyman limit is redshifted to redder wavelengths with increasing redshift, this makes an object become undetectable (`drop-out') from a given optical/near-infrared band. Several groups over the years have used this technique to identify high-$z$ star-forming galaxies, initially using optical data from the \HST\ Advanced Camera for Surveys (ACS) to select objects out to $z\sim 6$ \citep[e.g.][]{Steidel1999, Dickinson2004, Bunker2004}, and later exploiting near-infrared observations with the Wide Field Camera 3 (WFC3) to obtain large samples of $z \gtrsim 6$ galaxies \citep[e.g.][]{Wilkins2010, Bouwens2011, McLure2013, Schenker2013}.

Here we use the high-$z$ galaxy candidates selected with the dropout technique by \citet{Bouwens2015} in the \Hubble\ Ultra Deep Field (HUDF). Since we are interested in galaxies out to the highest redshifts, we only consider sources selected from the 4.7 \sqarcmin\ region in the HUDF with deep \HST/WFC3 near-infrared observations. We adopt the multi-band \HST\ catalogue of \citet{Illingworth2013}, built from a combination of all available \HST/ACS and WFC3 observations in the HUDF (for a complete list see table~2 of \citealt{Illingworth2013}), and refer to this as the `eXtreme Deep Field' (XDF) catalogue. 
The XDF catalogue includes observations in 9 bands, 5 in the optical, based on the ACS/WFC filters F435W, F606W, F775W, F814W and F850LP, and 4 in the near-infrared, taken with the WFC3 filters F105W, F125W, F140W and F160W. The $5\upsigma$ depth of a point source is $\magH \sim 29.8$, computed within a circular aperture of 0.35 arcsec diameter, while the $\txn{ACS}+\txn{WFC3}$ combined image used for the source extraction reaches a $5\upsigma$ depth of 31.2 within the same circular aperture \citep{Illingworth2013}. We do not use \Spitzer/IRAC data, since the vast majority of our galaxies are too faint to be detected in existing \Spitzer\ images. 

\begin{figure}
	\centering
	\resizebox{\hsize}{!}{\includegraphics{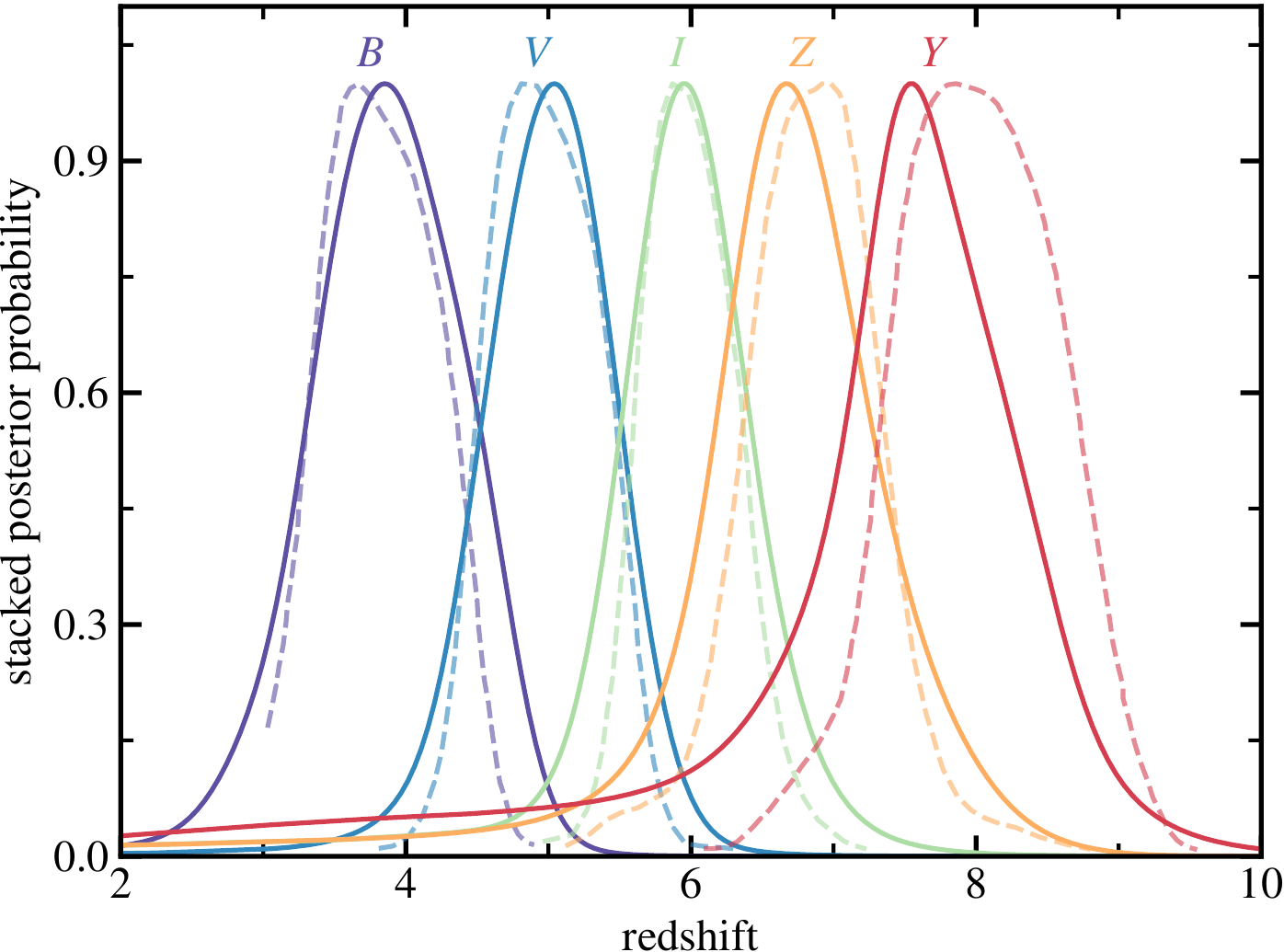}}
	\caption{Redshift distribution of the $B$, $V$, $I$, $Z$ and $Y$ dropouts, colour coded as in Fig.~\ref{fig:XDF_mag}. Dashed lines indicate the expected redshift distribution computed by \citet{Bouwens2015} by means of Monte Carlo simulations of artificial sources. Solid lines are computed by `stacking' (i.e., summing) the posterior probability distribution of redshift computed with the \beagle\ tool for each object, normalising the resulting distribution to a maximum value of 1, and then, following \citet{Bouwens2015}, convolving it with a Normal distribution with zero mean and standard deviation of 0.2.}
	\label{fig:stacked_redshift}
\end{figure}

We consider dropout galaxies in the filters $B_{435}$ ($z \sim 4$), $V_{606}$ ($z \sim 5$), $I_{775}$ ($z \sim 6$), $Z_{850}$ ($z \sim 7$) and $Y_{105}$ ($z \sim 8$), which we will indicate as $B$, $V$, $I$, $Z$ and $Y$ dropouts in the remainder of this paper. Details on the Lyman Break selection in each band can be found in sec~3.2 of  \citet{Bouwens2015} (see also their table~2). Fig.~\ref{fig:XDF_mag} shows the F160W magnitude distribution of galaxies selected in the different dropout bands. The vast majority ($\sim 80$ per cent) of the galaxies at $z\gtrsim 4.5$ ($V$ dropouts and above) have $\magH > 28$, while the 50 per cent completeness magnitude varies between a minimum of $\sim29.3$ (F105W filter, $I$ dropouts) to a maximum of $\sim29.7$ (F160W filter, $Y$ dropouts). The adopted Lyman Break selection produces, for each dropout band, the redshift distributions shown in Fig.~\ref{fig:stacked_redshift}. Galaxy redshifts are centred around the average values reported above, with Full Width Half Maximum (FWHM) of the redshift distribution set by the filter widths, i.e. $\delta z \sim 1$ for the $B$, $V$, $I$ and $Z$ dropouts, increasing to $\delta z \sim 2$ for the $Y$ dropouts. The absolute UV magnitude of the sources, computed from the median \magH\ magnitude in each dropout class, assuming a flat $f_\nu$ spectrum, and using the central redshift of each dropout band, varies from $\sim -19.1$ ($B$ dropouts), to $\sim -19.5$ ($V$), $\sim -19.65$ ($I$), $\sim -19.75$ ($Z$), $\sim -20.3$ ($Y$).

\subsection{Broad-band SED fitting of high-redshift galaxies in the HUDF}\label{sec:fitting_XDF}

\begin{table*}
	\centering
	\caption{Priors relative to the different free parameters used in the \beagle\ tool for the broad-band SED fitting of \HST/XDF photometry and for the full-spectrum fitting of \JWST/NIRSpec simulated spectra. The symbol $\mathcal{N}(\txn{mean};\, \txn{sigma})$ indicates a Gaussian (Normal) distribution.}
	\begin{tabular}{C{0.15\textwidth-2\tabcolsep} C{0.20\textwidth-2\tabcolsep} L{0.35\textwidth-2\tabcolsep}  C{0.125\textwidth-2\tabcolsep} C{0.125\textwidth-2\tabcolsep}}
\toprule


\multicolumn{1}{c}{Parameter}	    &  \multicolumn{1}{c}{Prior}  &  \multicolumn{1}{c}{Description} &  \multicolumn{1}{c}{XDF photometry} &  \multicolumn{1}{c}{NIRSpec spectra} \\     

\midrule

$z$						                & Uniform	$\in [0,15]$	   & Redshift & \ding{53} & \\

$\log (\M/\Msun)$		          & 	Uniform	$\in [5,12]$ 	 & Stellar mass (does not account for fraction of mass returned to the ISM by stellar mass loss) & \ding{53} &  \ding{53} \\

$\log (\tausfr / \txn{yr})$		& 	Uniform	$\in [7,10.5]$  & Time scale of star formation in a SFH described by a delayed exponential function. & \ding{53} &  \ding{53} 	\\

$\log (\t / \txn{yr})$		    &  Gaussian	$\mathcal{N}(8.5;\,0.5)$ truncated $\in [7,\tformm(z)]$    &  Age of onset of star formation in the galaxy	& \ding{53} &  \ding{53} \\

$\log (Z / \Zsun)$		        & 	Uniform	$\in [-2.2,0.25]$    &	Stellar metallicity & \ding{53} &  \ding{53}  \\

$\log (\Zism / \Zsun)$		    & 	Uniform	$\in [-2.2,0.25]$    &	Interstellar metallicity & & \ding{53}  \\

$\log (\sfrc / \MsunyrInv)$		    & 	Gaussian $\mathcal{N}(0;\,2)$ \; truncated $\in [-4,4]$   & Star formation rate averaged over the last $10^7$ yr	& \ding{53} &\ding{53}  \\

\logUs		                    & 	Uniform	$\in [-4,-1]$    &	 Effective gas ionization parameter & \ding{53} & \ding{53} \\

$\xid$		                    & 	Uniform	$\in [0.1,0.5]$    &	  Dust-to-metal mass ratio & \ding{53} & \ding{53} \\

$\tauV$		                    & 	Exponential $\exp(-\tauV)$ truncated $\in [0,5]$    &	 $V$-band attenuation optical depth & \ding{53} & \ding{53} \\

\mud		                      & 	Uniform	$\in [0,1]$    &	 Fraction of $V$-band attenuation arising in the diffuse ISM &  & \ding{53} \\

\bottomrule
	\end{tabular}
\label{tab:XDF_priors}	
\end{table*}

Similarly to \citet[][see their sec~4]{Chevallard2016}, we use the \beagle\ tool to fit the XDF photometry of 715 dropout galaxies with a self-consistent physical model that includes stellar emission, continuum+line emission from \Hii\ regions and diffuse ionized gas and dust attenuation. We do not model the emission from an AGN potentially contaminating the \HST\ photometry, as the expected number of $z>3$ type-1 AGNs in the 4.7 \sqarcmin\ field here considered is consistent with zero (e.g. see section~4.2.5 of \citealt{Grazian2015}). We let the redshift free to vary, and adopt a two-component star formation history constituted by a `smooth' function and a burst. The smooth component is described by a delayed exponential function $\psi(\tprime) \propto \tprime \exp{(-\tprime/\tausfr)}$, where \tausfr\ is the star formation timescale and \tprime\ the age of the galaxy, taken to lie between $10^7$ yr and the maximum time allowed since the onset of star formation at the galaxy redshift (see below). 
The burst (with constant star formation rate) covers the last $10^7\,\txn{yr}$ of star formation, the timescale over which $\sim 99.9$ per cent of the H-ionizing photons are emitted \citep[e.g.][]{Charlot1993, Binette1994}, and it is parametrised in terms of the `current star formation rate' \sfrc, i.e. the SFR averaged over the past 10 Myr. Decoupling the `past' star formation history and the `current' star formation rate allows us to obtain galaxy spectra with any contribution of emission lines relative to stellar continuum. We fix the maximum redshift for the formation of the first stars in a galaxy at $\zformm=15$, so that at any redshift $z$ the maximum allowed time since the onset of star formation is $\tformm(z) = \Tuniverse-\t(\zformm)$, where \Tuniverse\ and $\t(\zformm)$ refer to the age of the Universe at redshift $z$ and \zformm, respectively. We approximate the distribution of stellar metallicities in a galaxy, including the metallicity of young stars (with ages $\tprime \le 10^7$ yr, \Zyoung) and the interstellar metallicity (\Zism), with a single metallicity $Z$, i.e. $\Zyoung=\Zism=Z$.\footnote{The interstellar metallicity was indicated as \Zgas\ in the original paper describing the \beagle\ tool \citep{Chevallard2016}, while here we follow the nomenclature adopted in \citet{Gutkin2016}.} Following \citet[][see also \citealt{Charlot2001}]{Gutkin2016} we describe the properties of gas ionized by young stars by means of galaxy-wide (`effective') parameters. The ionization parameter \logUs\ determines the ratio of H-ionizing photons to gas density at the Str\"omgren radius of an effective star cluster, while the dust-to-metal mass ratio \xid\ (`depletion factor') sets the amount of depletion of heavy elements onto dust grains (see sec~2.3 of \citealt{Gutkin2016} for a discussion of depletion factors).\footnote{Our definition of \logUs\ implies a volume-averaged ionization parameter $\langle U \rangle = 9/4 \, \Us$ (see equation~1 of \citealt{Hirschmann2017}).} We model the effect of dust attenuation on stellar and gas emission by appealing to the two-component model (diffuse ISM + birth clouds) of \citet{Charlot2000}, parametrized in terms of the total attenuation optical depth \tauV\ and the fraction of attenuation arising in the diffuse ISM $\mud$. We account for the effect of absorption from the intergalactic medium (IGM) by means of the average prescription of \citet{Inoue2014}.

Following the Bayesian approach adopted in \beagle, we define the posterior probability distribution of the model free parameters \thetab\ as
\begin{equation}\label{eq:posterior}
\conditional{\thetab}{\Db,H} \propto \pi(\thetab) \, \mathcal{L}(\thetab) \, ,
\end{equation}
where \Db\ indicates the data, $H$ the adopted model, $\pi(\thetab)$ the prior distribution and $ \mathcal{L}(\thetab)$ the likelihood function. We adopt independent priors for all parameters, uniform for the parameters $z$, $\log (\M/\Msun)$, $\log (\tausfr / \txn{yr})$, $\log (Z / \Zsun)$, \logUs\ and $\xid$, Gaussian for $\log (\t / \txn{yr})$ and $\log (\sfrc / \MsunyrInv)$ and exponential for \tauV\ (see Table~\ref{tab:XDF_priors}).\footnote{We note that the adopted prior for \tauV\ would correspond to the non-informative, scale-invariant Jeffreys prior for a one-dimensional Gaussian likelihood function depending only on \tauV.} We consider a multi-variate Gaussian likelihood function with independent errors $\upsigma_i$ on each measurement $y_i$
\begin{equation}\label{eq:likelihood}
- 2\, \ln \mathcal{L}(\thetab^k) = \sum_i \left [ \frac{y_i-\hat{y}_i(\thetab^k)}{\upsigma_i} \right ]^2 \, ,
\end{equation}
where the summation index $i$ runs over all observed bands (i.e., bands with positive errors, and negative or positive fluxes), $y_i$ is the observed flux, $\upsigma_i = \sqrt{\upsigmao^2+(\upsigma_0\,y_i)^2}$, where \upsigmao\ is the observational error and $\upsigma_0 = 0.02$ is an additional error term that we add to avoid obtaining results dominated by systematic uncertainties (see sec.~4.2 of \citealt{Chevallard2016}) and $\hat{y_i}(\bmath{\thetab}^k)$ indicates the fluxes predicted by our model for a set of parameters $\bmath{\thetab}^k$. 

We adopt the Nested Sampling algorithm \citep{Skilling2006} as implemented in \multinest\ \citep{Feroz2009} to sample the posterior probability distribution of the 9 free parameters $\thetab_\textsc{xdf} = [z, \logM, \tausfr, \t, Z, \sfrc, \logUs, \xid, \tauV]$. It is worth briefly pausing to discuss the potential risk of over-fitting our data by using such a flexible physical model. The 9 photometric bands used in the fitting cover the wavelength range $0.4 \lesssim \lambda/\micron \lesssim 1.6$, hence they mainly probe the rest-frame UV emission of $z\gtrsim3$ galaxies, and only bands redward the Lyman break (4 bands for the $Y$ dropouts) provide constraints on the galaxy physical properties. In star-forming galaxies, the UV continuum emission is mostly sensitive to the recent ($\lesssim 10^8$ yr) star formation history and to dust attenuation, while other parameters, such as the mass of older stars and the physical conditions of gas, have little influence on the emission at these wavelengths. Since in this work we aim at simulating NIRSpec observations covering a large variety of intrinsic galaxy spectra, extending beyond those observed in relatively bright galaxies at $z\lesssim 4$, we must also account for the variation of model parameters largely unconstrained by existing photometric observations. For this reason we adopt the flexible, 9-parameters model described above, and combine the weak constraints provided by \HST\ photometry on some model parameters with well established relations among galaxy physical quantities to obtain physically-motivated combinations of parameters (see Section~\ref{sec:spectra_XDF}). We then validate this approach by comparing our model predictions with external (photometric and spectroscopic) data-sets at redshift \range{z}{2}{8} (Section~\ref{sec:external_constraints}).

The results of the \beagle\ fitting of the XDF data are summarised in Fig.~\ref{fig:stacked_redshift} in terms of the stacked posterior probability distribution of redshift. This is computed by combining the redshift probability distribution of each galaxy using a kernel-density estimator. 
Similarly to an histogram, Fig.~\ref{fig:stacked_redshift} represents a density plot, where the solid curves depend on both the density of galaxies at each redshift and on the redshift probability distribution of each individual source. Fig.~\ref{fig:stacked_redshift} provides an interesting visual comparison among the redshift distributions obtained by \citet{Bouwens2015} from the analysis of Monte Carlo simulations of artificial sources (see their section~4.1) and those computed in this work. Given the very different approaches adopted to compute the two sets of lines of Fig.~\ref{fig:stacked_redshift}, the figure highlights a good agreement between the two methods. The differences among the two approaches are larger for the $Y$ dropouts, for which the peaks of the two distributions are separated by ($\delta z\sim 0.5$), possibly because of the low constraining power of data for the $Y$ dropouts (low \SN, few bands red-ward the continuum break) combined with a low number of sources ($\txn{N}_Y =  34$).

\subsection{Linking model spectra and observed galaxies in the  HUDF}\label{sec:spectra_XDF}

The \beagle\ tool provides us with the posterior probability distribution of the 9 free parameters of the model (reported in column 5 of Table~\ref{tab:XDF_priors}), along with sets of selected observables (e.g., full spectrum, absolute and apparent magnitudes) and derived quantities (e.g. rates of H- and He-ionizing photons). These are obtained by sampling the posterior probability distribution of model parameters using the \multinest\ algorithm (see section~3.3 of \citealt{Chevallard2016} for more information on the \beagle\ output). From a purely statistical perspective, we could associate with a given XDF source any spectrum from the set drawn using \multinest\ in this way. However, since the data are consistent with a broad range of model spectra, and a correspondingly broad range of model parameters, this approach would not ensure that the model spectra match other, independent observables (e.g. galaxy colours at longer wavelengths, emission line measurements), nor that the selected combinations of model parameters are consistent with measured relations among galaxy physical quantities (e.g. the mass-metallicity relation). Obtaining a physically-motivated distribution of emission line strengths is particularly relevant for XDF galaxies, which exhibit faint continuum emission and will likely be detected only via emission lines with \JWST/NIRSpec. By obtaining such a physically-motivated distribution of emission lines, we can produce simulated NIRSpec observations with realistic \SN\ ratios. This, in turn, should ensure that the constraints on the galaxy physical parameters derived in Section~\ref{sec:nirspec_simulations} can be applied to future `deep' NIRSpec observations of \HST-selected high-redshift galaxies.  
 
After some experimentation, we find that requiring for the fitted galaxies to follow three well-established relations between galaxy physical parameters produces observables in agreement with different types of observations. These relations include a relation between (i) stellar mass and star formation rate (`main sequence' of star-forming galaxies); (ii) stellar mass, star formation rate and gas-phase metallicity (`fundamental' metallicity relation); (iii) interstellar metallicity and ionization parameter. We adopt the (redshift-dependent) relation between stellar mass and star formation rate of \citet{Speagle2014}, which they derive from a broad compilation of stellar mass and star formation rate measurements at redshift \range{z}{0}{6}. Their relation (see their equation~28) can be expressed as  
\begin{equation}\label{eq:speagle}
\begin{split}
\log(\sfrc/\MsunyrInv) = & [0.84 - 0.026 \, (\tprime/\Gyr)] \, \log(\Mstar/\Msun) - \\
& [6.51 - 0.11 \, (\tprime/\Gyr)] \, ,
\end{split}
\end{equation}
where $\tprime$ indicates the age of the Universe at redshift $z$.

Following \citet{Williams2018}, we consider the relation between stellar mass, star formation rate and gas-phase metallicity of \citet{Hunt2016}, which they obtain from the analysis of the UV-to-infrared photometry and optical emission lines of a sample of $\sim1000$ star-forming galaxies at redshift \range{z}{0}{3.7}. The relation of \citet{Hunt2016} can be expressed as  
\begin{equation}\label{eq:hunt}
\begin{split}
\logOH  = & -0.14 \log{(\sfrc/\MsunyrInv)} \\
& + 0.37 \log{(\Mstar/\Msun)} + 4.82  \, .
\end{split}
\end{equation}

Finally we adopt the relation between interstellar metallicity and ionization parameter of \citet{Carton2017}, which they derive by adopting the `theoretical' dependence of ionization parameter on interstellar metallicity of \citet{Dopita2006} ($\logUs \propto 0.8  \log{\Zism}$), and computing the zero-point of this relation by using metallicities and ionization parameters of SDSS DR7 galaxies estimated by \citet{Brinchmann2004}. The relation between \logUs\ and \Zism\ can then be expressed as
\begin{equation}\label{eq:carton}
\begin{split}
\logUs = -0.8 \log{(\Zism/\Zsun)} -3.58 \, .
\end{split}
\end{equation}

In practice, we enforce these relations by multiplying the posterior probability distribution of equation~\eqref{eq:posterior} by three `weight' functions which depends on the above parameters. In order to more densely populate the tails of these relations, we adopt a Student's-t weight function with 3 degrees of freedom, which is significantly broader than a Gaussian function. The Student's-t function is expressed as
\begin{equation}\label{eq:student}
f(x) = \frac{6\sqrt{3}}{\pi \left( 3+x^2 \right )^2} \, \frac{1}{\upsigma_x},
\end{equation}
where $x$ is the standardised variable 
\begin{equation}
x = \frac{x^\prime - \overbar{x^\prime}}{\upsigma_x} \, ,
\end{equation}
and $x^\prime$ indicates the value of the model parameter in the posterior distribution computed with \beagle\ and $\overbar{x^\prime}$ the value of the parameter predicted by equations~\eqref{eq:speagle}--\eqref{eq:carton}. We fix the scatter to $\upsigma_x=0.3$ for the mass--star formation rate relation (see, e.g., \citealt{Speagle2014, Shivaei2015}), and to $\upsigma_x=0.2$ for the `fundamental' metallicity relation and for the metallicity--ionization parameter relation.
Equations~\eqref{eq:speagle}--\eqref{eq:student} provide the conditional distribution of a model parameter (star formation rate, gas-phase metallicity, ionization parameter) given other model parameters, for example equations~\eqref{eq:speagle} and \eqref{eq:student} define the conditional probability \conditional{\sfrc}{\Mstar,z}. This conditional distribution does not depend on the data \Db, hence it can be readily interpreted as a conditional prior distribution linking different model parameters.

\begin{figure}
	\centering
	\resizebox{\hsize}{!}{\includegraphics{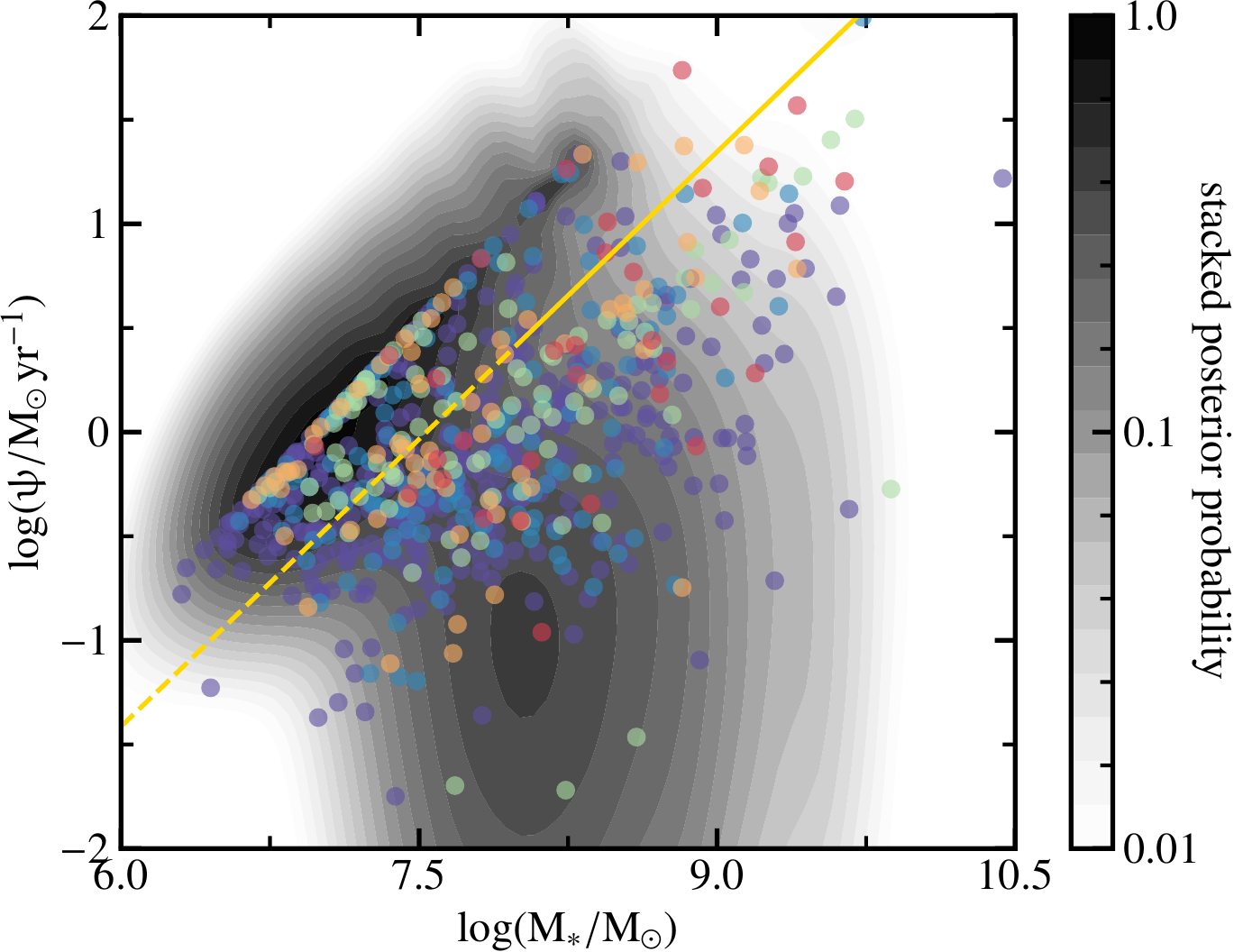}}
	\caption{Stacked two-dimensional posterior probability distribution of mass and star formation rate for the $B$, $V$, $I$, $Z$ and $Y$ dropouts (grey density contours, plotted on a logarithmic scale). Circles, colour coded as in Fig.~\ref{fig:XDF_mag}, indicate the pairs [\Mstar, \sfrc] corresponding to a random realisation of model parameters drawn from the re-weighted posterior probability distribution. The solid yellow line indicates the mass--star formation rate relation measured by \citet{Santini2017} at \range{z}{5}{6} from $\sim50$ galaxies selected from 4 gravitationally-lensed \HST\ Frontier Fields, while the dashed line is an extrapolation of this relation at lower stellar masses.}
	\label{fig:stacked_mass_sfr}
\end{figure} 

It is well known that the photometric degeneracy between the Lyman and Balmer breaks can make a galaxy photometric SED be equally well described by two different models with widely different redshifts \citep[e.g.][]{Ilbert2006}. As the \beagle\ tool can identify these multiple redshift solutions, in this work we restrict the analysis to redshifts consistent with the expected dropout redshift of \citet{Bouwens2015}. To achieve this, among the combinations of model parameters \thetab\ obtained with \beagle, we only consider those satisfying the condition $z>z_\txn{min}$, where $z_\txn{min}=2$ for the $B$ dropouts, $z_\txn{min}=3$ for the $V$, $z_\txn{min}=4$ for the $I$ and $z_\txn{min}=5$ for the $Z$ and $Y$ ones (Fig.~\ref{fig:stacked_redshift}). We then randomly draw, for each galaxy, a set of parameters among those sampled by \multinest, where the probability of drawing any set is proportional to the re-weighted posterior probability distribution. We can repeat the random draw (with replacement) $N$ times to obtain $N$ model spectra consistent with the observed \HST\ photometry, and corresponding to combinations of $[\Mstar, \sfrc]$ satisfying the adopted mass-star formation rate relation. This is particularly relevant to evaluate selection effects and to study the consequences of deriving population-wide relations among physical parameters (e.g. mass-metallicity, mass-star formation rate) from a finite number of observations. Both applications will be part of successive works, while in the remainder of this paper we consider a single random draw to associate a model spectra to each XDF source, and refer to the ensemble of 715 model spectra as a `mock catalogue'.

\begin{figure*}
	\centering
	\resizebox{\hsize}{!}{\includegraphics{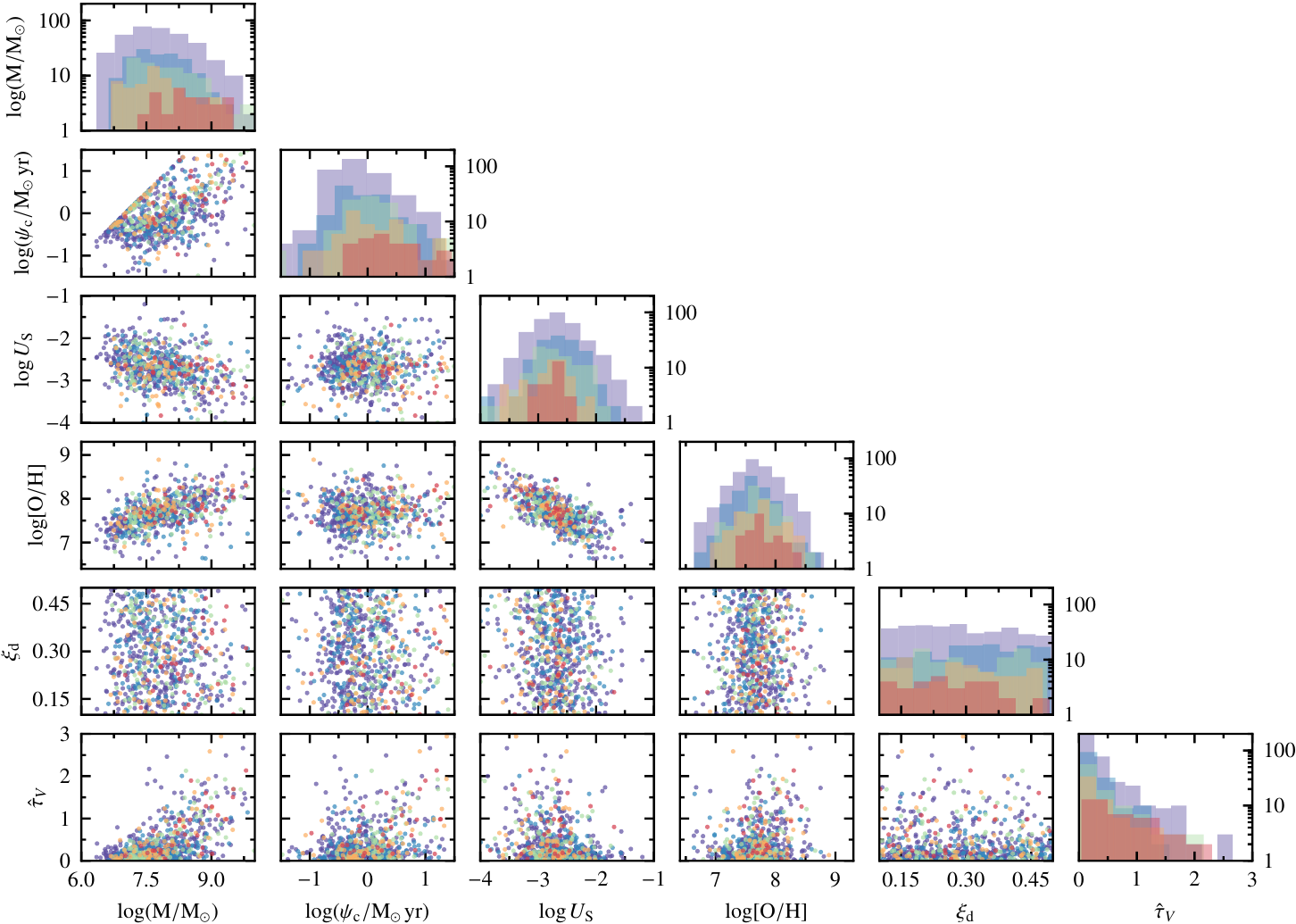}}
	\caption{Distribution of model free parameters of the mock catalogue of galaxy spectra computed as detailed in Sections~\ref{sec:fitting_XDF}--\ref{sec:spectra_XDF}. The colour coding indicates mock galaxies at different redshifts, and it is the same as in Fig.~\ref{fig:XDF_mag}. The off-diagonal panels show the relation between each pair of parameter $[\logM, \sfrc, \logUs, \logOH, \xid, \tauV]$, while the diagonal panels display, by means of histograms, the distribution of each parameter.}
	\label{fig:mock_input}
\end{figure*}

We show in Fig.~\ref{fig:stacked_mass_sfr} by means of (grey) density contours on a logarithmic scale the two-dimensional stacked posterior probability distribution of stellar mass \Mstar\ and star formation rate \sfrc, along with the pairs $[\Mstar, \sfrc]$ drawn from the re-weighted posterior probability distribution (circles of different colours). By analogy with the one-dimensional stacked posterior probability distribution of redshift (see Section~\ref{sec:fitting_XDF}), we compute the two-dimensional stacked posterior probability distribution of \Mstar\ and \sfrc\ summing the individual probability distributions computed with a kernel-density estimator. Fig.~\ref{fig:stacked_mass_sfr} shows the effect of varying in our model both the current star formation rate and total stellar mass when only rest-frame UV observations are available. The observed \HST\ photometry of a large fraction of XDF galaxies can be reproduced equally well by our model for widely different combinations of \Mstar\ and \sfrc, which makes the stacked distribution occupy a broad region of Fig.~\ref{fig:stacked_mass_sfr}. 
We note that the sharp upper envelope created by the coloured circles in Fig.~\ref{fig:stacked_mass_sfr} is caused by the fixed duration (10 Myr) of the current burst of constant SF, which limits the maximum \sfrc\ attainable at fixed \Mstar. 


Fig.~\ref{fig:mock_input} shows the distribution of the main model free parameters of the mock catalogue. The low constraining power of \HST\ photometry with respect to most model parameters makes these distributions reflect the adopted priors, i.e. Gaussian for \sfrc, exponential for \tauV\ and uniform for the other parameters (see Table~\ref{tab:XDF_priors}). Also, no strong correlations are visible except for the relation between $\log \Mstar$ and $\log \sfrc$ that we enforced. 
It is worth briefly discussing the absence in Fig.~\ref{fig:mock_input} of a correlation between mass and metallicity, and metallicity and ionization parameter. The mass--metallicity relation is well-established at low redshift (e.g. from SDSS data, \citealt{Tremonti2004}), while observations targeting galaxy populations similar to those used in our analysis, i.e. faint high-redshift galaxies, depict a less clear picture. Recent results from the VUDS survey suggest that highly star-forming galaxies with stellar masses comparable to those of our mock galaxies (\range{\log \M/\Mstar}{6.5}{9.5}) span a broad range of gas-phase metallicities \range{\logOH}{7.5}{8.5} at fixed stellar mass \citep{Calabro2017}. Similar results have been obtained by \citet{Izotov2015} and \citet{Jimmy2015} from the analysis of low-mass local `analogues' of high redshift galaxies. These findings can have a physical origin, since the metal content of star-forming gas results from the complex interaction between enrichment from previous stellar generations, AGN and stellar feedback and gas accretion, or can be driven by systematic uncertainties in the metallicity estimators \citep[e.g.][]{Kewley2008}. 

Similarly, an anti-correlation between gas-phase metallicity and ionization parameter has been observed at low redshift \citep{Carton2017}, and can be expected on theoretical grounds \citep[e.g.][]{Dopita2006}, but the redshift evolution and dispersion of the relation for galaxy populations as those considered in this work are largely unknown.
While future observations may show that certain combinations of model parameters (e.g. high metallicity and large ionization parameter) are not well suited to describe the conditions of photoionized gas in high-redshift star-forming galaxies, we have decided not to impose any relation between these parameters in our mock galaxy catalogue. This will enable users to adopt any relation between mass--metallicity and metallicity--ionization parameter to select sub-samples of sources from the different realisations of the mock catalogue.

\subsection{Comparison of model spectra with independent data}\label{sec:external_constraints}

\begin{figure*}
	\captionsetup[subfigure]{width=0.9\textwidth}
	\centering
	\begin{subfigure}{.47\hsize}
		\resizebox{\hsize}{!}{\includegraphics{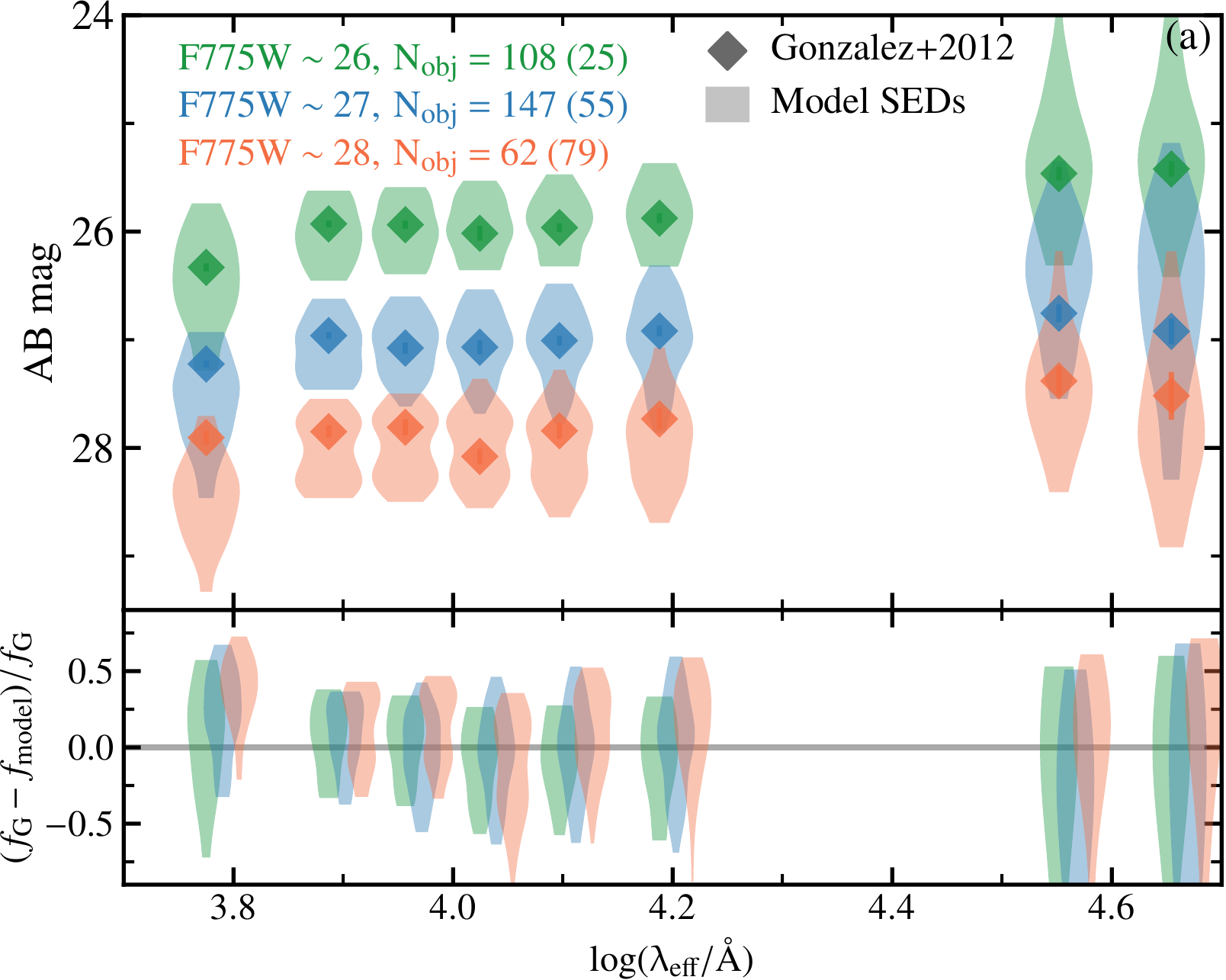}}
		\label{fig:gonzalez_a}
	\end{subfigure}
	\hspace{14pt}
	\begin{subfigure}{.47\hsize}
		\resizebox{\hsize}{!}{\includegraphics{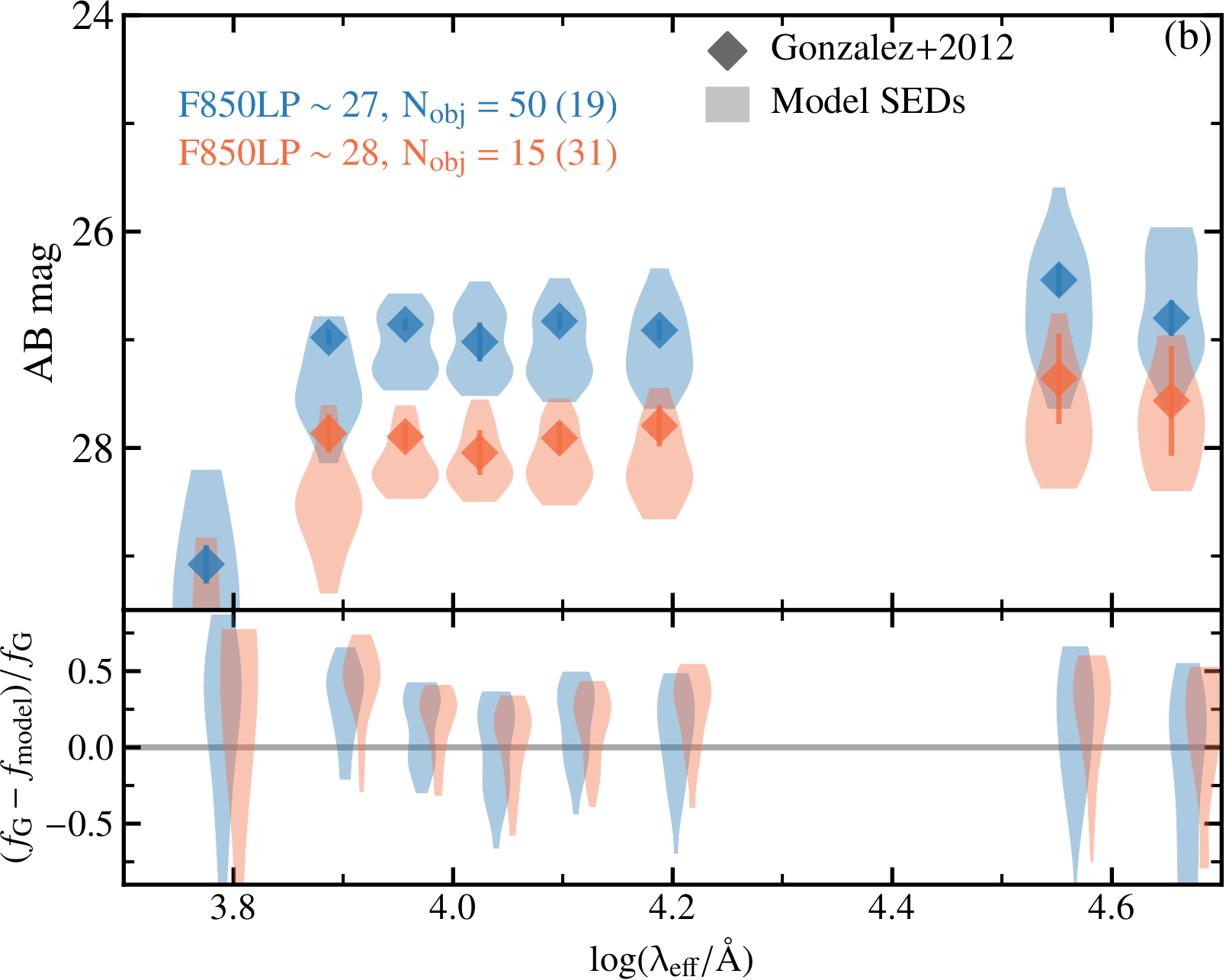}}
		\label{fig:gonzalez_b}
	\end{subfigure}
	\caption{(a) In the top panel, diamonds with ($1\,\upsigma$) error bars indicate the stacked photometric SEDs of redshift $\sim4$ galaxies ($B$ dropout) of \citet{Gonzalez2012} in three bins of observed F775W magnitudes, used as a proxy for the rest-frame UV luminosity. Shaded `violins' indicate the magnitude distribution (95 per cent interval) covered by our mock spectra based on $B$ dropouts in the XDF, split in the same UV luminosity bins as in \citet{Gonzalez2012}. We indicate in the inset legend the number of objects in each \citet{Gonzalez2012} stack, and in parenthesis the number of objects in our mock catalogue entering each bin. In the bottom panel we show the distribution of the flux differences between the \citet{Gonzalez2012} stacks and our mock spectra. For each band, the shaded regions of different colours are slightly shifted horizontally for clarity. (b) Same as (a), but for $V$ dropout galaxies (redshift $\sim5$). We did not plot the median SED for the $\txn{F850W}\sim26$ bin because of the low (4) number of galaxies in our mock catalogue falling in the bin.}
	\label{fig:gonzalez}
\end{figure*}

Our approach to build a catalogue of mock galaxy spectra guarantees, by construction, that these match the \HST\ photometry of XDF dropouts. In this section, we compare the mock spectra with other observables, namely near infrared photometry from the Infrared Array Camera (IRAC) onboard the \Spitzer\ \textit{Space Telescope}, and measurements of the \CIII\ emission line performed with different instruments. The lower sensitivity of \Spitzer\ with respect to \HST\ only enables the detection in the IRAC band 1 (3.6 \micron\ filter, \IRACone) and band 2 (4.5 \micron\ filter, \IRACtwo) of a minority of XDF dropouts with $\magH \lesssim 27$ (see Fig.~\ref{fig:XDF_mag}). For this reason, we have to rely on \Spitzer\ observations of stacked galaxies and of individual (brighter) galaxies over a much wider area than the XDF region.

\begin{figure}
	\centering
	\resizebox{\hsize}{!}{\includegraphics{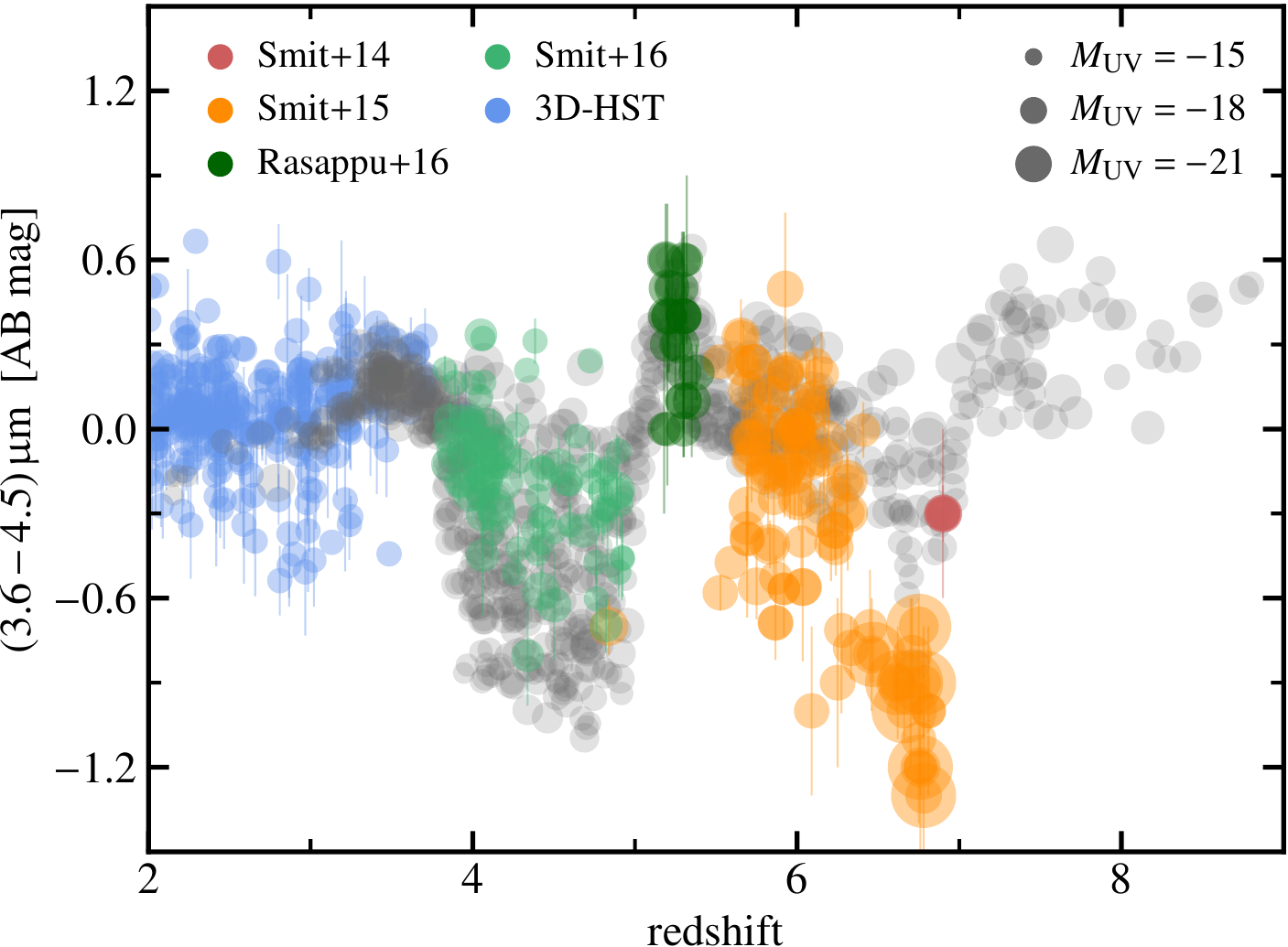}}
	\caption{Comparison of the \Spitzer/IRAC colour $3.6 \micron - 4.5 \micron$ at $2\lesssim z \lesssim 8$ from the literature (coloured circles) with that of the galaxies in our mock catalogue (grey circles). In order of increasing redshift, we show galaxies with spectroscopic redshifts at $2 \le \zspec \le 4$ from 3D-HST/GOODS-North and GOODS-South \citep[cyan circles, ][]{Skelton2014}, objects from the spectroscopic sample of \citet{Smit2016} (light green) and from the photometric samples of \citet{Rasappu2016} (dark green), \citet{Smit2015} (orange) and \citet{Smit2014} (red). We only plot objects for which the quoted uncertainty on the IRAC colour is $\le 0.30$. For clarity, we only show error bars when these are $\ge 0.05$. As indicated in the inset legend, the area of the circles is proportional to the absolute UV magnitude of each galaxy \Muv.}
	\label{fig:IRAC_colours}
\end{figure} 

\begin{figure}
	\centering
	\resizebox{\hsize}{!}{\includegraphics{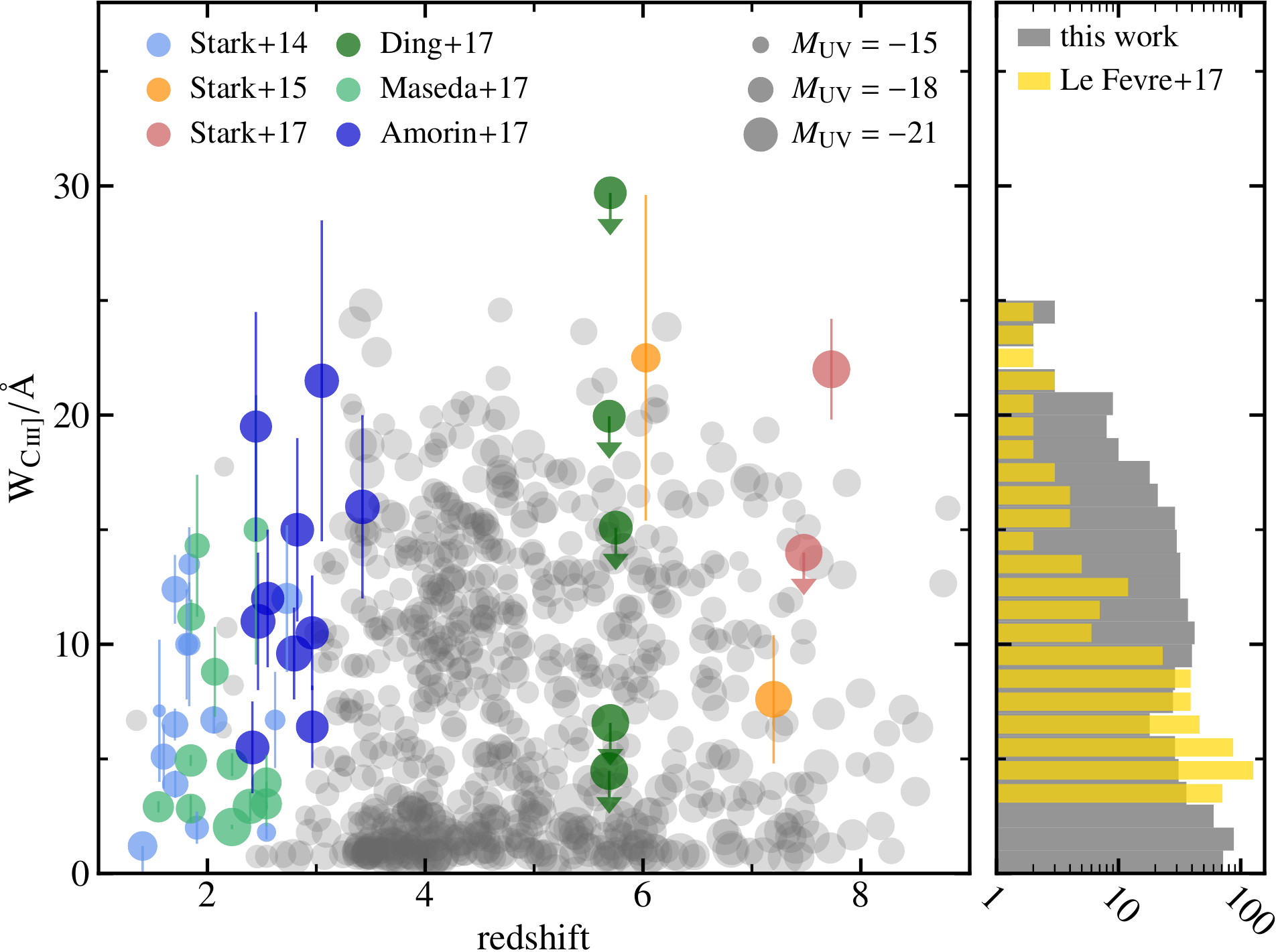}}
	\caption{Left: comparison of the \CIII\ (rest-frame) equivalent widths \EWCIII\ from literature (coloured circles) with those measured from the spectra of our mock catalogue (grey circles). Literature values are taken, in order of increasing redshift, from \citet[cyan circles]{Stark2014}, \citet[light green]{Maseda2017}, \citet[blue]{Amorin2017}, \citet[dark green]{Ding2017}, \citet[orange]{Stark2015} and \citet[red]{Stark2017}. The upper limits indicate 3 $\upsigma$ limits. As in Fig.~\ref{fig:IRAC_colours}, the area of the circles is proportional to the absolute UV magnitude \Muv\ of each galaxy. Right: distribution of \CIII\ equivalent widths in our mock catalogue and in the sample of \range{z}{2}{3.8} star-forming galaxies of \citet{LeFevre2017}. As  \citet{LeFevre2017} consider only galaxies with $\textnormal{W}_{\textnormal{C\textsc{iii}}]}/\AA \ge 3$, we renormalize the distribution of \citet{LeFevre2017} to the total number of objects in our sample with $\textnormal{W}_{\textnormal{C\textsc{iii}}]}/\AA \ge 3$.}
	\label{fig:CIII}
\end{figure}

We compare in Fig.~\ref{fig:gonzalez} the predicted optical-to-near infrared SEDs ($\HST+\Spitzer$ bands) of our mock catalogue with the stacked SEDs computed by \citet{Gonzalez2012} at $z\sim4$ and $z\sim5$ galaxies. \citet{Gonzalez2012} consider \HST/ACS, \HST/WFC3 and \Spitzer/IRAC observations in the GOODS-South region \citep{Giavalisco2004, Bouwens2012}, and compute the median SEDs of $B$, $V$, $I$ and $Z$ dropouts in bins of observed UV luminosity. In practice, they compute the median \HST\ fluxes of the sources in each UV luminosity bin, and perform aperture photometry on the median-combined image in each IRAC band. They evaluate the uncertainties on the median fluxes through bootstrap resampling. 
We note that when comparing model predictions with \Spitzer/IRAC fluxes of $z\gtrsim4$ galaxies, one must consider the effect of optical emission lines contaminating the \IRACone\ and \IRACtwo\ bands. At \range{z}{3}{5} ($B$ dropouts), \HaNIISII\ contaminates the \IRACone\ band, while at $z\gtrsim 5$ both IRAC bands can be contaminated by either \HbOIII\ or \HaNIISII. This makes the IRAC colours computed from the stacked SEDs of $z\gtrsim 5$ galaxies highly dependent on the emission line properties and redshift distribution of the sources entering the stacks, and this dependence is exacerbated in stacks computed from a low number of sources (e.g. see fig.~7 of \citealt{Gonzalez2012}). For this reason, we only consider the $z\sim4$ and $z\sim5$ stacked SEDs of \citet{Gonzalez2012} computed from a minimum of 15 individual sources.
Fig.~\ref{fig:gonzalez}(a) and (b) show a good agreement among the SEDs of our mock catalogue and the median SEDs \citet{Gonzalez2012} in all the UV luminosity bins considered. This test is particularly important since the IRAC fluxes, unlike the \HST\ ones, are not matched to any observation during the mock catalogue creation. Fig.~\ref{fig:gonzalez} hence demonstrates that, on average, the mass and ages of evolved stars in our mock galaxies, traced by the IRAC fluxes probing the rest-frame optical-to-near infrared emission of galaxies, are consistent with the observed values.

The \Spitzer/IRAC colour \IRACcolour\ can be used to further test the agreement between observations and mock spectra. As we noted above, the redshift evolution of this colour is sensitive to the presence of optical emission lines, especially the two groups \HbOIII\ and \HaNIISII, contaminating the \IRACone\ and \IRACtwo\ bands (e.g. observationally, \citealt{Stark2013, deBarros2014}; theoretically, \citealt{Wilkins2013}). We therefore compare in Fig.~\ref{fig:IRAC_colours} the evolution of the \IRACcolour\ colour in the redshift range $2\lesssim z \lesssim 8$ predicted by our mock catalogue with data from different sources. These include, from low to high redshift, observations from 3D-HST \citep{Skelton2014}, from the spectroscopic sample of \citet{Smit2016} and from the photometric samples of \citet{Rasappu2016}, \citet{Smit2015} and \citet{Smit2014}. Since our mock catalogue is based on observations from the small, 4.7 \sqarcmin\ XDF area, while data are drawn from much larger areas ($\gtrsim100$ \sqarcmin), we encode in the circle sizes the rest-frame UV luminosity \Muv\ of each object: the larger the circle, the more luminous the galaxy. 
Fig.~\ref{fig:IRAC_colours} highlights the ability of the mock spectra to reproduce the sharp IRAC colour change at $z\sim 3.8$ ($z\sim5$) caused by the entry of \Ha\ in the \IRACone\ (\IRACtwo) band, including the most extreme blue and red colours, which are caused by $\txn{EW}(\HaNIISII)\sim 1000$. We note that the different extremes reached by the IRAC colours at \range{z}{3.8}{5} ($\IRACcolour \sim -1$) and at \range{z}{5}{6}  ($\IRACcolour \sim 0.8$) are likely caused by the different widths of the IRAC filters, $\sim0.75$ \micron\ for \IRACone\ filter and $\sim 1$ \micron\ for \IRACtwo\ one. This difference translates into a different contamination of the integrated broad-band flux of each band, at fixed emission line equivalent width (\EW). 
At redshifts \range{z}{5.5}{6.6} the \IRACone\ band is contaminated by the group of lines \HbOIII\ and the \IRACtwo\ band by \HaNIISII, therefore the \IRACcolour\ colour depends on the relative intensities of H-Balmer lines \emph{vs} \OIII, i.e. on the physical conditions of ionized gas, especially metallicity and ionization parameter. 
In the narrow redshift window \range{z}{6.6}{6.9}, the \IRACtwo\ band is free of strong emission lines, while \IRACone\ is contaminated by \HbOIII. \citet{Smit2015} exploit this property to select extreme emission lines galaxies with accurate photometric redshifts. They search for such extreme objects in the 5 CANDELS fields ($\sim 900 \; \sqarcmin$), and find $\sim 20$ sources with $\IRACcolour\lesssim -1$. These are relatively bright galaxies falling in a narrow redshift range, hence, not surprisingly, they do not appear in our mock catalogue, which reaches at most $\IRACcolour \sim -0.6$ at \range{z}{6.6}{6.9}. Overall, Fig.~\ref{fig:IRAC_colours} indicates that the strengths of the \HaNIISII\ and \HbOIII\ lines in our mock catalogue are consistent with those inferred from \Spitzer/IRAC colours of \range{z}{2}{8} galaxies. This validation is particularly important since the strength of the optical emission lines in our mock catalogue will translate into a distribution of \SN\ in the NIRSpec simulations (see Section~\ref{sec:simulation_characteristics}), therefore directly affecting the constraints on galaxy physical parameters form NIRSpec spectra discussed in Section~\ref{sec:constraints}. 

The above tests only provide indirect constraints on spectroscopic features through the contamination of broad band filters by emission lines. As we have already noted, existing observatories do not allow the detection of optical emission lines at $z\gtrsim4$, hence to study the evolution of galaxy spectral features across the widest redshift range one must rely on UV emission lines. The \Lya\ line is the most luminous emission line at UV wavelengths, but radiative transfer effects caused by its resonant nature make the comparison of model predictions and observations non trivial \citep[e.g.,][]{Verhamme2006}. We therefore consider another bright UV line, the doubly-ionized carbon line \CIII, which has been observed both at low \citep[e.g.][]{Berg2016, Senchyna2017} and high redshift \citep[e.g.][]{Erb2010, Stark2015}. We compare in Fig.~\ref{fig:CIII} the redshift evolution of the \CIII\ equivalent width from literature with that predicted by our mock catalogue. In the same figure, we also show with a histogram the distribution of \CIII\ equivalent widths in our mock catalogue, and the distribution derived by \citet{LeFevre2017} from observations at redshifts \range{z}{2}{3.8}.  
Fig.~\ref{fig:CIII} indicates that while most galaxies in our catalogue exhibit $\EWCIII \lesssim 5$ \AA, a significant number of mock spectra attain substantially larger \EWCIII\ values, reaching the most extreme $\EWCIII \gtrsim 20$ \AA\ found by \citet{Stark2015, Stark2017} and \citet{Amorin2017}. Interestingly, Fig.~\ref{fig:CIII} also reveals a larger fraction of galaxies with \CIII\ equivalent widths $\gtrsim 10$ \AA\ in our catalogue relative to that in the sample of \citet{LeFevre2017}. This is likely a consequence of the higher redshifts and lower luminosities of our mock galaxies with respect to the sample of \citet{LeFevre2017} ($z\lesssim 4$, $i_\txn{AB} \le 25$), which imply, on average, younger and more metal poor stellar populations, able to power stronger \CIII\ emission.

The comparisons of our mock spectra with external observables presented in this section validate our semi-empirical approach to build a mock galaxy catalogue based on \HST\ photometry in the HUDF. In particular, the above tests demonstrate the ability of our catalogue to match the rest-frame optical continuum emission of $z\sim4$ and $z\sim5$ galaxies (Fig.~\ref{fig:gonzalez}), the contamination of the strongest optical emission lines to \Spitzer/IRAC bands (Fig.~\ref{fig:IRAC_colours}), and the observed range of \CIII\ equivalent widths at \range{z}{1}{8} (Fig.~\ref{fig:CIII}).
 
\section{Simulating and fitting \JWST/NIRSpec observations}\label{sec:nirspec_simulations}

In this study, we focus on simulations of deep (100~ks) observations performed with \JWST/NIRSpec. We consider the multi-object spectroscopy (MOS) mode and adopt the low spectral resolution configuration NIRSpec/prism (PRISM/CLEAR spectral configuration), which enables the simultaneous observation of up to $\sim 200$ objects over the full spectral range \range{\lambda}{0.6}{5.3} \micron.



\subsection{Simulating \JWST/NIRSpec spectra}\label{sec:ETC_simulations}

\begin{figure}
	\centering
	\resizebox{\hsize}{!}{\includegraphics{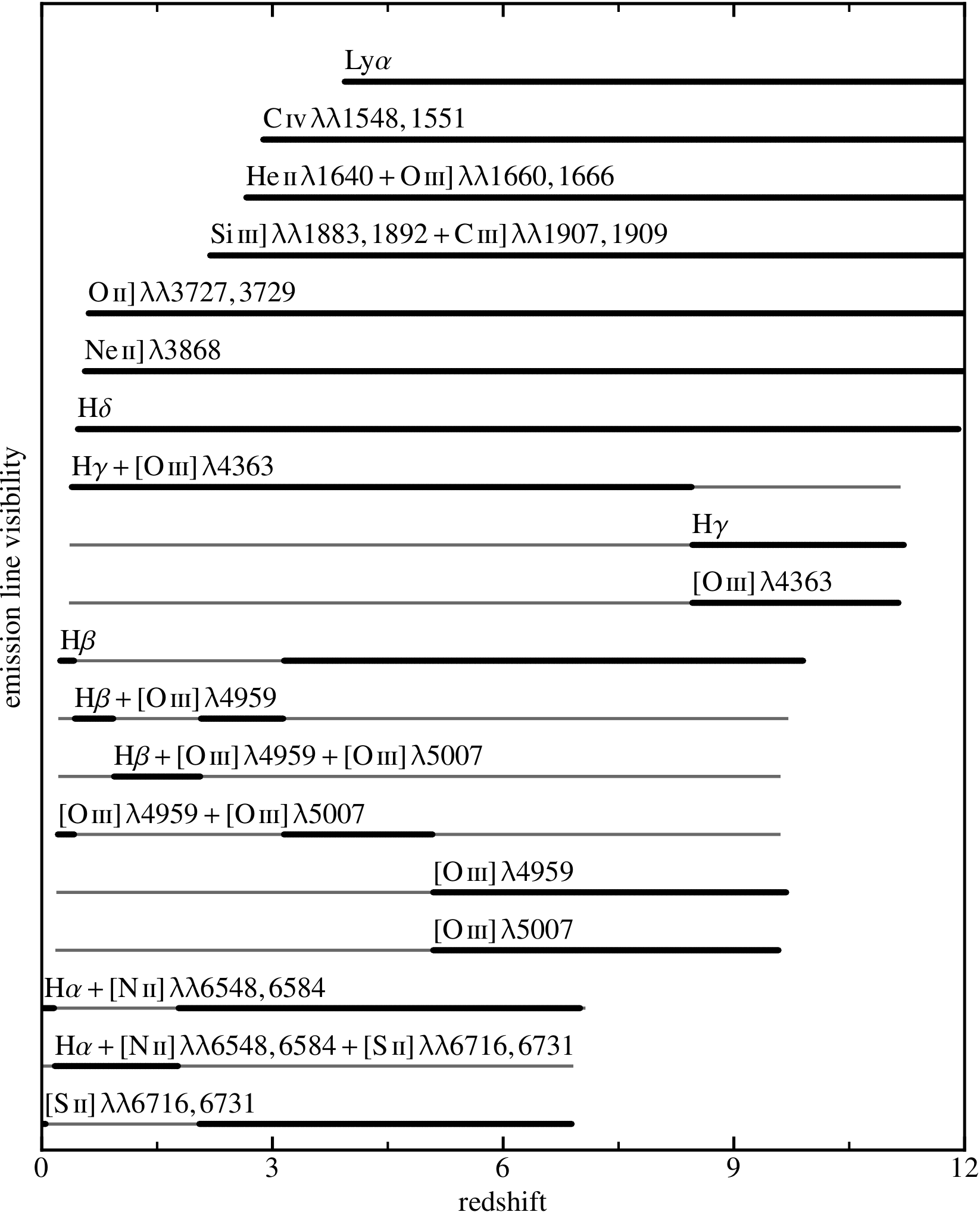}}
	\caption{Visibility and blending/unblending of the strongest UV and optical emission lines as a function of redshift, for NIRSpec/prism observations. The thin grey lines indicate the lines visibility as a function of redshift, computed assuming the nominal NIRSpec wavelength coverage $\lambda = 0.6 - 5.3 \, \micron$. The thick black lines show the unblended single lines or blended group of lines which will be observable at different redshifts.}
	\label{fig:EL_visibility}
\end{figure} 

A detailed description of the approach adopted to simulate NIRSpec spectra can be found in Appendix~\ref{app:etc_simul}. Here, we only provide an overview of our approach and highlight a few key features of the simulations.
For each galaxy, we start from the (noiseless) mock spectrum obtained using the procedure described in Section~\ref{sec:semi_empirical}. This high-resolution spectrum is redshifted to the photometric redshift obtained with the \beagle\ analysis and then rebinned to the spectral pixel size of the NIRSpec/prism configuration. We then use an idealized model accounting for the response of the telescope and of the instrument to derive the number of electrons per-second generated at detector level in each pixel. Combined with a noise model, this allows us to generate a simulated (noisy) spectrum. This approach is very similar to the one used by many exposure-time calculators (ETCs), but simpler than the elaborate two-dimensional scheme used by \JWST's official ETC `Pandeia' \citep{Pontoppidan2016}.


In our simulations, we also account for the fact that a significant fraction of the object light may fall outside of the standard 3-shutter slitlet, especially for extended objects (the projected size of the aperture of an individual micro-shutter is $0.20 \times 0.42$ arcsec). In practice, we compute the (wavelength-dependent) throughput of an extended source with effective radius \re, described as an exponential profile with S\'ersic index 1.5 \citep{Shibuya2015}, and average over all source positions within the aperture of the central shutter of the slitlet. Averaging over random ellipticities and position angles produces small corrections ($\lesssim$ few per cent), which we therefore ignore. Following the approach of \citet{Williams2018} (see their section~5.2), we associate an effective radius to each galaxy in the catalogue by adopting the (redshift-dependent) relation between galaxy UV luminosity and galaxy size of \citet{Shibuya2015}
\begin{equation}
\re = r_\txn{e}^0 \left ( \frac{L_\textsc{uv}}{L_\textsc{uv}^0} \right )^{0.27} \, ,
\end{equation}
where $r_\txn{e}^0 = 6.9 \times (1 + z) - 1.2$ kpc and $L_\textsc{uv}^0$ is the UV luminosity corresponding to a UV magnitude $M_\textsc{uv}^0 = -21 $. This leads to median sizes of 0.9 (0.13), 0.75 (0.12), 0.63 (0.11), 0.54 (0.10) and 0.48 (0.10) kpc (arcsec) for the $B$, $V$, $I$, $Z$, $Y$ dropouts, respectively. The total throughput at $\lambda = 2.5 \, \micron$ corresponding to the median sizes above is between 34 per cent ($B$ dropouts) and 38 per cent ($Y$ dropouts) for the aperture and light profile adopted.

 
The resolution $R=\lambda/\Delta \lambda$ of the NIRSpec prism has a strong wavelength dependence, from a minimum of $R\sim 30$ at $\lambda=1.2$ \micron\ to a maximum of $R\sim 300$ at $\lambda=5$ \micron. This makes groups of neighbouring emission lines appear blended or unblended depending on a galaxy redshift. Since UV and optical emission lines will provide the firmest constraints on the physical parameters of faint galaxies at high redshift, we have summarised in Fig.~\ref{fig:EL_visibility} the visibility and blending/unblending of different lines as a function of redshift. Emission lines in Fig.~\ref{fig:EL_visibility} are considered unblended when their redshifted wavelengths are separated by a $\Delta \lambda$ corresponding to 2.2 detector pixels, i.e. the typical instrumental spectral resolution element size of an extended source which uniformly fills the micro-shutter aperture of width $\sim 0.2$ arcsec (along the spectral direction). We note that the ETC simulations performed here do not account for the impact of the object size on the spectral response, implying that emission lines always fall into a single detector pixel (see Appendix~\ref{app:e_rates}). This approximation has a minor impact on the results of our analysis, as the only strong lines potentially affected are the two components of \OIII, whose ratio is fixed by atomic physics, and the \Ha\ and \NII\ lines. The low metallicity of our galaxies and absence of an AGN component in our models imply that \NII\ is in any way too weak to provide any constraint on the physical properties of mock galaxies. Fig.~\ref{fig:EL_visibility} indicates that groups of emission lines in the UV such as $\HeII+\OIIIuv$ and $\SiIII+\CIII$ are blended at all redshifts, while at $z \gtrsim 4$ \Hb\ is unblended with respect to \OIIIa\ and \OIIIb, and these [O\,{\sc iii}] lines are unblended at $z\gtrsim 5$. The \Ha\ line is blended with \NII\ at $2 \lesssim z \lesssim 6.5$, while \Hg\ is blended with \mbox{[O\,{\sc iii]}\lam 4636} up to $z\sim8$. Note also that all the lines in Fig.~\ref{fig:EL_visibility} but $\CIIIa+\CIIIb$ and $\OIIa+\OIIb$ are unblended when using the $R\sim 1000$ disperser, while the $R\sim 2700$ disperser allows one to resolve also those doublets at most redshifts. 
Fig.~\ref{fig:EL_visibility} also highlights the ability of NIRSpec/prism observations to provide the simultaneous measurement of the main UV and optical emission lines for galaxies at \range{z}{3}{10}: \Ha\ is accessible up to $z\sim 7$, \Hb\ to $z \sim 10$ and \OIII\ to $z \sim 9.5$.

\begin{figure*}
	\centering
	\begin{subfigure}{.47\hsize}
		\resizebox{\hsize}{!}{\includegraphics{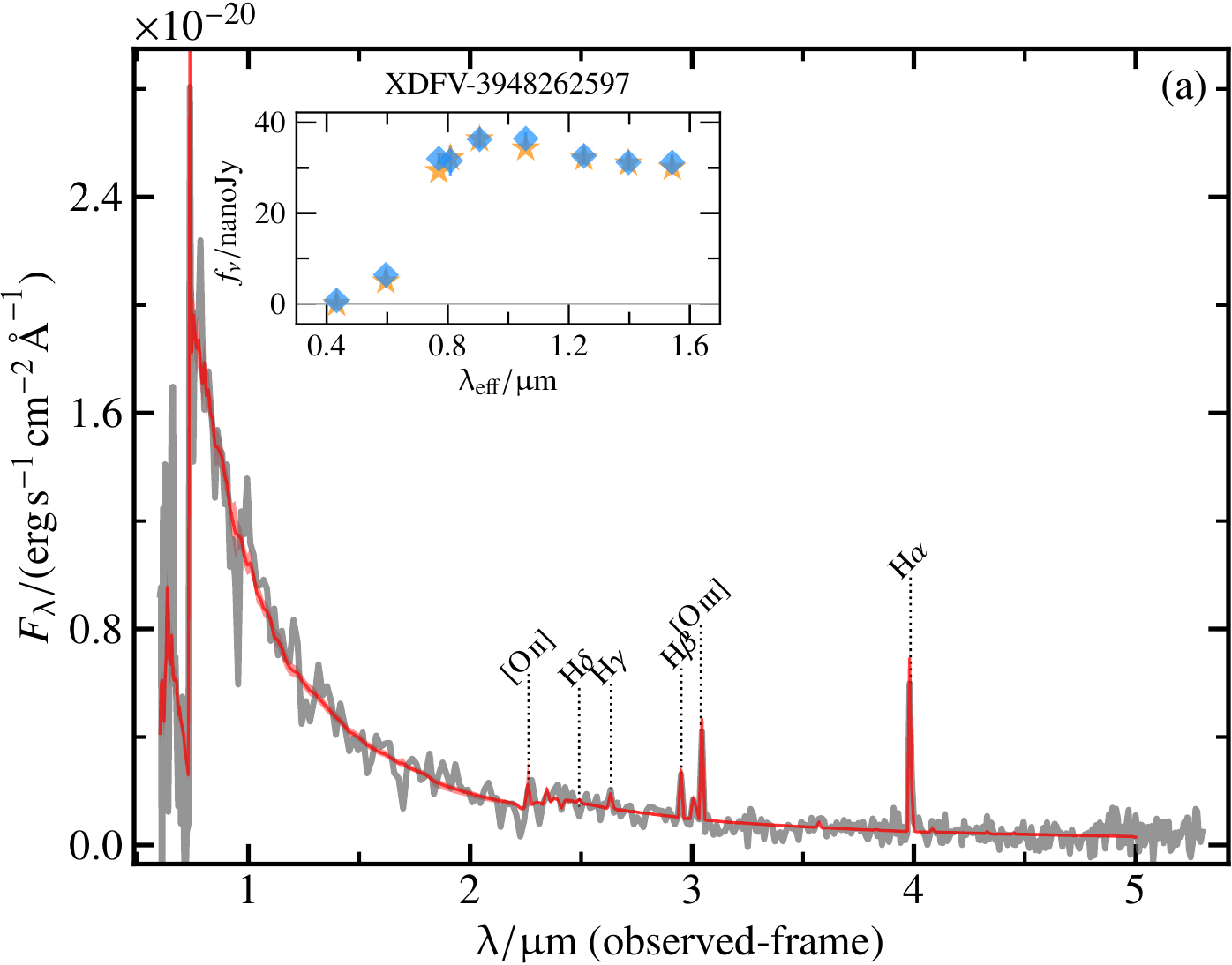}}
	\end{subfigure}
	\begin{subfigure}{.47\hsize}
		\resizebox{\hsize}{!}{\includegraphics{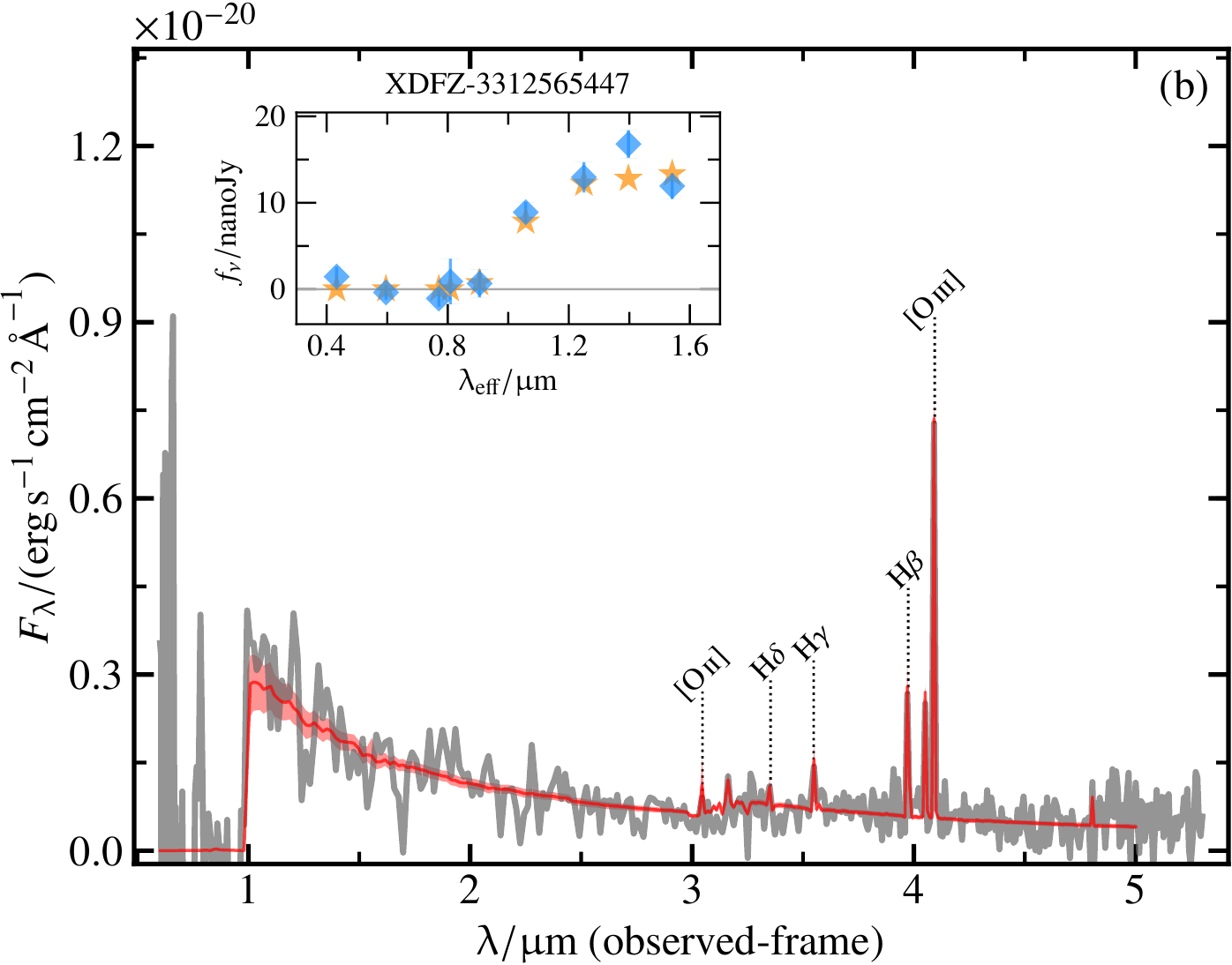}}
	\end{subfigure}
	\caption{(a) Simulated NIRSpec/prism spectrum (thick grey line, corresponding to a $\sim100$ ks exposure) of a $V$ dropout (XDFV-3948262597, $\magH=27.70$) placed at redshift $z=5.072$, with stellar mass $\log({\Mstar/\Msun}) \sim 8.18$, star formation rate $\sfrc/\MsunyrInv \sim 0.7$ and specific star formation rate $\log{(\ssfr/\yrInv)} \sim -8.4$. The red line and red shaded region indicate the posterior median and 95 per cent credible interval, respectively, from the \beagle\ fit of the simulated spectrum. (b) same as (a), but for a $Z$ dropout (XDFZ-3312565447, $\magH=28.74$), with $z=7.176$, $\log{(\Mstar/\Msun)} \sim 8.9$, $\sfrc/\MsunyrInv \sim 5.4$ and $\log{(\ssfr/\yrInv)} \sim -8.3$. The small inset in each panel shows the observed XDF photometry (blue diamonds, from \citealt{Bouwens2015}) along with the model photometry predicted by \beagle\ for the set of physical parameters associated to these sources with the procedure outlined in Sec.~\ref{sec:spectra_XDF} (orange stars).}
	\label{fig:nirspec_spectra}
\end{figure*} 

\begin{figure*}
	\centering
	\resizebox{.7\hsize}{!}{\includegraphics{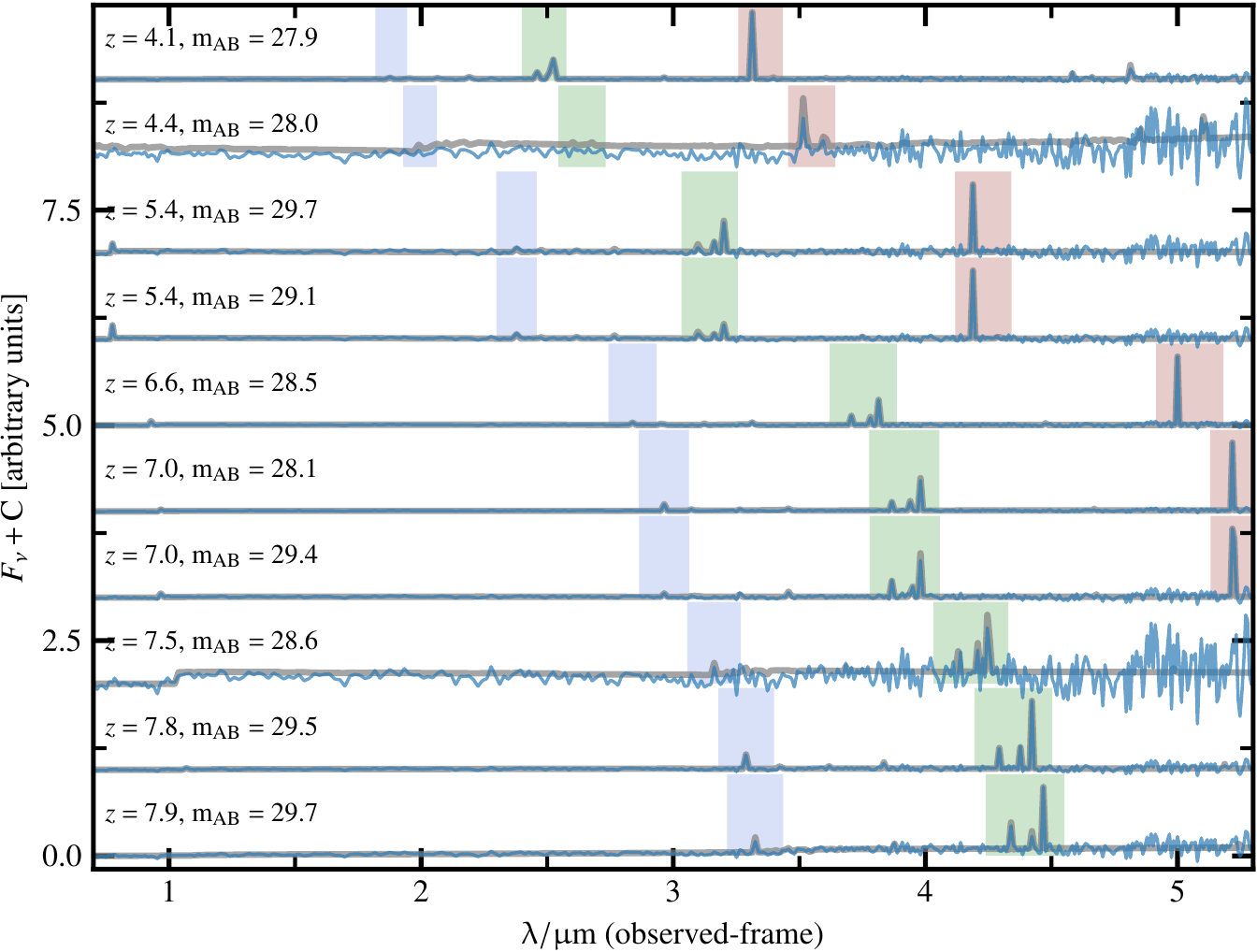}}
	\caption{Illustration of simulated NIRSpec/prism spectra (corresponding to a $\sim100$ ks exposure) of 10 galaxies at \range{z}{4}{8}. The grey line indicates the noiseless input spectrum, while the blue line shows the simulated one. Each spectrum is normalized to its maximum value, then shifted vertically for clarity. The magnitudes reported in the figure are observed ones and refer to the \HST/WFC3 F160W filter, while the redshift is the \beagle-based photo-$z$ of the mock galaxy. Shaded regions indicate the location of the main optical emission lines \OIInoL\ (blue), \HbOIII\ (green) and \HaNIISII\ (red).}
	\label{fig:simulated_spectra}
\end{figure*}

We show in Fig.~\ref{fig:nirspec_spectra} and \ref{fig:simulated_spectra} a few examples of simulated NIRSpec/prism spectra. Fig.~\ref{fig:nirspec_spectra}(a) shows a $z=5.072$ galaxy with $\magH=27.70$,  and Fig.~\ref{fig:nirspec_spectra}(b) one at $z=7.176$, with $\magH=28.74$. 
Fig.~\ref{fig:nirspec_spectra}(a) shows a high signal-to-noise detection ($\SNHb\sim8$) of the Balmer lines \Ha\ and \Hb, and of the Oxygen lines \OIIIa\ and \OIIIb. Fig.~\ref{fig:nirspec_spectra}(b) illustrates that the sensitivity of \JWST/NIRSpec will allow us to obtain highly significant detections ($\SNHb\sim10$) of emission lines (\Hg, \Hb, \OIIIa\ and \OIIIb) even for a fainter galaxy at redshift $z\sim 7$. The inset of Fig.~\ref{fig:nirspec_spectra}(b) shows a flux excess in the \HST\ F140W band, but no strong emission line that can cause such an excess falls in the F140W band at $z\sim7$, making the origin of the observed flux excess unclear.
Fig.~\ref{fig:simulated_spectra} provides an overview of NIRSpec/prism observations of 10 galaxies at \range{z}{4}{8} covering a broad range of continuum luminosities \range{\magH}{27.9}{29.7}. The figure highlights how the large star formation rates per unit stellar mass and young ages of high redshift galaxies should enable statistically significant detections of emission lines even in $\magH \gtrsim 29$ galaxies out to $z\sim9$.

\subsection{Equivalent width and \SN\ distribution of emission lines}\label{sec:simulation_characteristics}

\begin{figure}
	\centering
	\resizebox{\hsize}{!}{\includegraphics{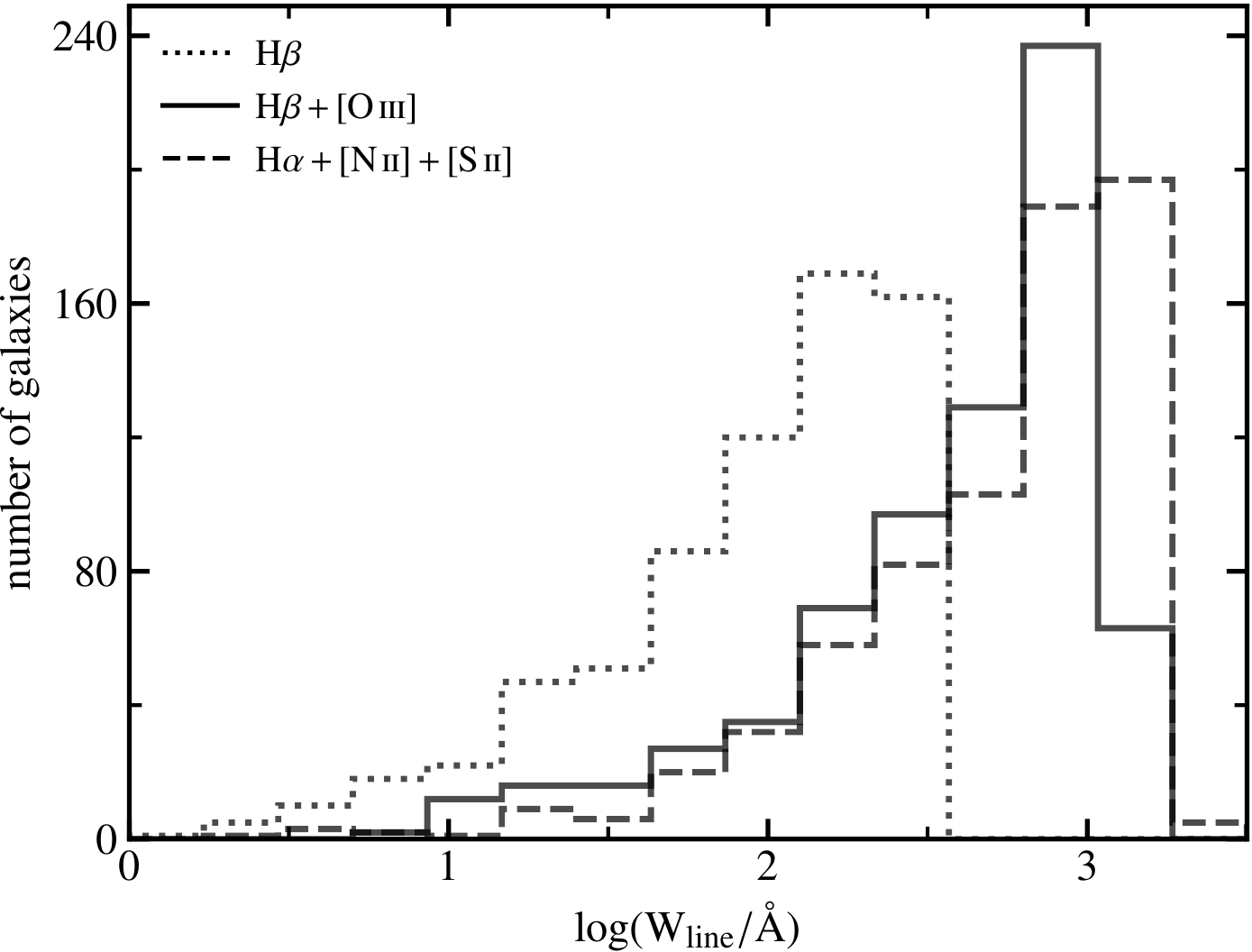}}
	\caption{Distribution of equivalent widths of \Hb\ (short-dashed line), $\Hb+\OIII$ (solid line) and $\Ha+\NII+\SII$ (long-dashed line), in our simulated galaxies.}
	\label{fig:EW_distrib}
\end{figure} 

We show in Fig.~\ref{fig:EW_distrib} the distribution of equivalent widths of \Hb\ and of the two groups of lines \HbOIII\ and \HaNIISII\ computed from the mock spectra. The mean (median) redshift of the galaxies in the mock catalogue is $z\sim4.8$ ($\sim4.5$), and the mean (median) equivalent width of $\HaNIISII \sim 450 \, \txn{\AA}$ ($\sim600 \, \txn{\AA}$), which compares favourably with the value $ \sim 400 \, \txn{\AA}$ found by \citet{Smit2016} at redshift \range{z}{3.8}{5.0}, and of $\sim 550 \, \txn{\AA}$ and $\sim 600 \, \txn{\AA}$ found by \citet{Rasappu2016} at redshift $5.1 \lesssim z \lesssim 5.4$ in their photometric and spectroscopic samples, respectively. 

\begin{figure}
	\centering
	\resizebox{\hsize}{!}{\includegraphics{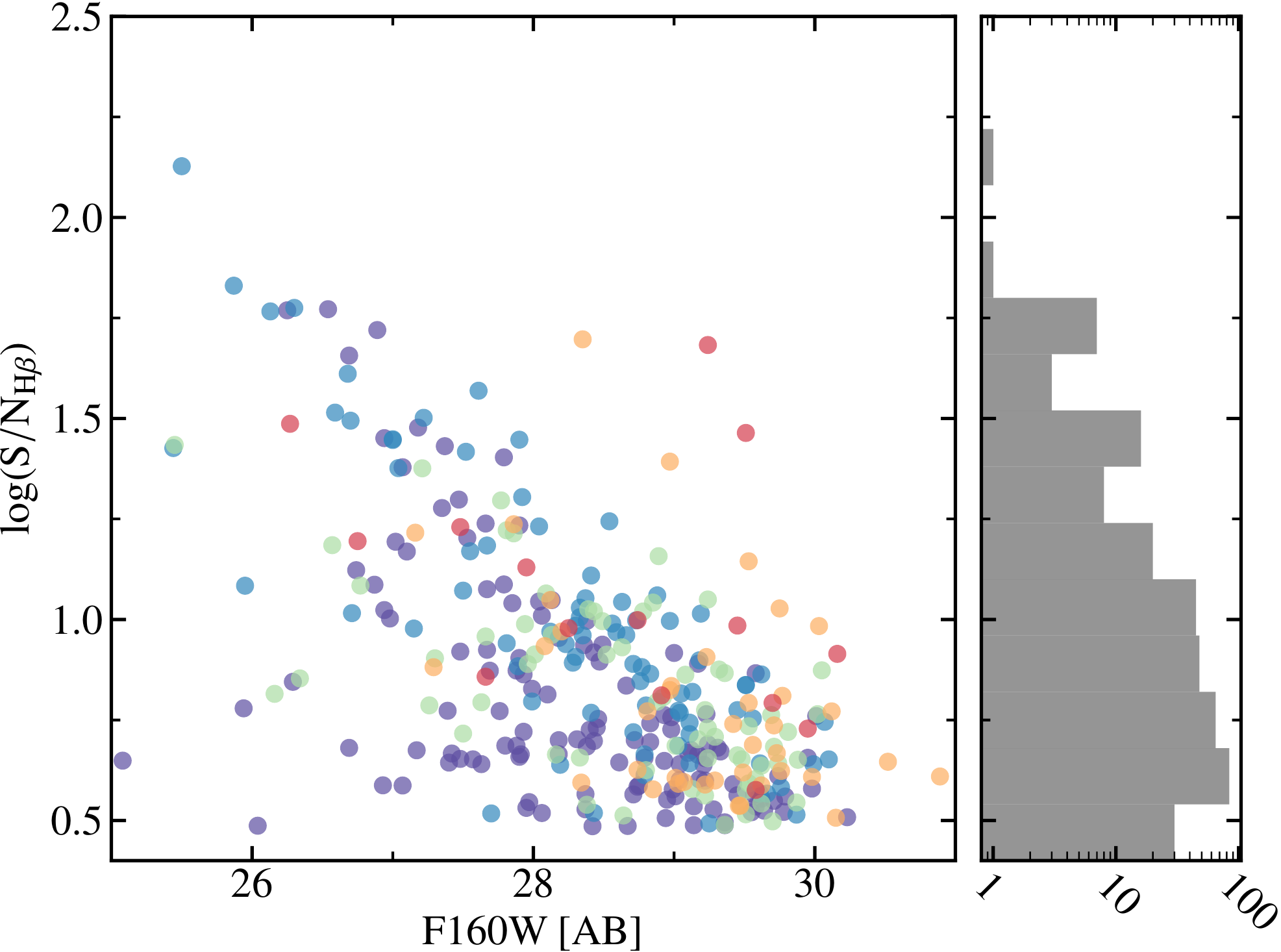}}
	\caption{Relation between S/N of the \Hb\ emission line and observed F160W magnitude. The colour coding is the same as in Fig~\ref{fig:XDF_mag} and reflects galaxies in different dropout classes.}
	\label{fig:S_to_N_mag}
	

\end{figure}

Fig.~\ref{fig:S_to_N_mag} shows the relation between the signal-to-noise ratio of the \Hb\ line \SNHb\ and the observed F160W magnitude for all galaxies in the mock catalogue, while Table~\ref{tab:Hb_SN} indicates the fraction of galaxies in bins of F160W magnitude with $\SNHb> 3$, 5 and 10. The \SN\ reported in Fig.~\ref{fig:S_to_N_mag} and Table~\ref{tab:Hb_SN} accounts for the effect of the instrumental spectral response, while our simulations do not account for this effect. The consequence is, on average, a $\sim35$ per cent larger emission line \SN\ in our simulated spectra with respect to a computation accounting for the spectral response (see Appendix~\ref{app:etc_simul}).

\begin{table} 
	\centering
	\caption{Fraction of objects in different bins of observed F160W magnitude with $\SNHb > 3$, 5 and 10 (in parentheses the number of objects in each magnitude bin).}
	\vspace{0pt}
	\begin{tabular}{C{0.40\columnwidth-2\tabcolsep} C{0.20\columnwidth-2\tabcolsep} C{0.20\columnwidth-2\tabcolsep}  C{0.20\columnwidth-2\tabcolsep} }
\toprule

\multicolumn{1}{c}{\magH}	    &  \multicolumn{1}{c}{$\SN \ge 3$}  & \multicolumn{1}{c}{$\SN \ge 5$}  &  \multicolumn{1}{c}{$\SN \ge 10$}   \\     

\midrule

$m \le 28$	 (199) & 0.51 & 0.42 & 0.29 \\

$ 28.0 < m \le 28.5$ (78)	  & 0.65 & 0.49 & 0.17 \\

$ 28.5 < m \le 29.0$ (104)	  & 0.49 & 0.33 & 0.07 \\

$ 29.0 < m \le 29.5$ (160)	  & 0.39 & 0.14 & 0.02 \\

$ 29.5 < m \le 30$ (124)	  & 0.40 & 0.13 & 0.02 \\

$ m > 30$ (44)	  & 0.27 & 0.16 & 0.00 \\

\bottomrule
\end{tabular}
\label{tab:Hb_SN}	
\end{table} 

Fig.~\ref{fig:S_to_N_mag} and Table~\ref{tab:Hb_SN} indicate that NIRSpec/prism `deep' observations will detect most $z\gtrsim4$ galaxies in the HUDF through their emission lines, with typical \range{\SNHb}{3}{30}. We expect more than 40 per cent of the XDF dropouts with $\magH < 28.5$ to exhibit $\SNHb>5$, and to detect with $\SNHb>3$ about one third of the faintest ($\magH > 30$) galaxies. The reported \SNHb\ values imply an even larger \SN\ of \Ha\ (visible out to $z\sim 7$), and likely of \OIIIb\ (out to $z\sim 9.5$), given the large $\OIIInoL/\Hb$ ratios observed at high redshift \citep[e.g.][]{Strom2017}.\footnote{Detector noise dominates the noise budget of NIRSpec observations of faint targets at the wavelengths relevant for this analysis. This causes a nearly linear increase of the line \SN\ with increasing received flux, at fixed observed wavelength (see also Appendix~\ref{app:etc_simul}). Assuming a constant photon conversion efficiency of the instrument, standard emission line intensities $\txn{I}_{\scriptsize \Ha}$, $\txn{I}_{\scriptsize \Hb}$ and $\txn{I}_{\scriptsize \OIIInoL}$ \citep{Steidel2016}, and ignoring the effect of dust attenuation, the \SN\ distribution in Fig.~\ref{fig:S_to_N_mag} would increase by a factor $\txn{I}_{\scriptsize \Ha}/\txn{I}_{\scriptsize \Hb} \cdot \lambda_{\scriptsize \Ha}/\lambda_{\scriptsize \Hb} \sim 3.89$ for \Ha\ and $\txn{I}_{\scriptsize \OIIInoL}/\txn{I}_{\scriptsize \Hb} \cdot \lambda_{\scriptsize \OIIInoL}/\lambda_{\scriptsize \Hb} \sim 4.36$ for \OIIIb. The factors $\lambda_{\scriptsize \Ha}/\lambda_{\scriptsize \Hb}$ and $\lambda_{\scriptsize \OIIInoL}/\lambda_{\scriptsize \Hb}$ account for the different number of photons, and hence of photo-electrons, generated at different wavelengths at fixed emitted energy.} As we will discuss in Section~\ref{sec:constraints}, the simultaneous detection at high \SN\ of several hydrogen and metal emission lines enables tight constraints on several key galaxy properties, such as star formation rates and gas-phase metallicities.



\subsection{Spectroscopic redshifts determination}\label{sec:spec_z}

\begin{table} 
	\centering
	\caption{Fraction of objects in different bins of redshift and observed F160W magnitude for which the deep NIRSpec/PRISM observations considered in this work would allow the detection of at least 1 or 2 emission lines with $\SN \ge 3$ and $\SN \ge 5$. The number of objects in each bin is given in parentheses.}
	\vspace{0pt}
	\begin{tabular}{C{0.19\columnwidth-2\tabcolsep} C{0.25\columnwidth-2\tabcolsep} C{0.14\columnwidth-2\tabcolsep}  C{0.14\columnwidth-2\tabcolsep} C{0.14\columnwidth-2\tabcolsep}  C{0.14\columnwidth-2\tabcolsep}}
\toprule

\multirow{2}{=}{\centering Redshift}   &  \multirow{2}{=}{\centering \magH}    & \multicolumn{2}{c}{$\SN \ge 3$}  & \multicolumn{2}{c}{$\SN \ge 5$}    \\     

	\cmidrule(lr){3-4} \cmidrule(lr){5-6} 
	
	&	&	$\Nlines \ge 1$	& $\Nlines \ge 2$	 & $\Nlines \ge 1$	 & $\Nlines \ge 2$	\\

\midrule

\multirow{4}{=}[-16pt]{\centering $z \le 5$}	&     $m \le 28$ \nObjBin{147} & 0.98 & 0.95  & 0.95 & 0.90  \\

 & $ 28 < m \le 29$ \nObjBin{108}	  & 0.99 & 0.94  & 0.93 & 0.80 \\

& $ 29 < m \le 30$ \nObjBin{158}	  & 0.96 & 0.84 & 0.91 & 0.68  \\

& $ m > 30$ \nObjBin{19}	  & 0.84 & 0.68 & 0.68 & 0.26 \\

\addlinespace[8pt]

\multirow{4}{=}[-16pt]{\centering $5 < z \le 7$}	&     $m \le 28$ \nObjBin{45} & 0.96 & 0.89 & 0.87 & 0.87 \\

 & $ 28 < m \le 29$ \nObjBin{62}	  & 1.00 & 0.97  & 0.93 & 0.87 \\

& $ 29 < m \le 30$ \nObjBin{98}	  & 0.95 & 0.87 & 0.83 & 0.66 \\

& $ m > 30$ \nObjBin{20}	  & 0.95 & 0.75 & 0.70 & 0.40 \\

\addlinespace[8pt]

\multirow{4}{=}[-16pt]{\centering $z >7$}	&     $m \le 28$ \nObjBin{7} & 1.00 & 1.00 & 1.00 & 0.70 \\

 & $ 28 < m \le 29$ \nObjBin{12}	  & 0.83 & 0.75 & 0.75 & 0.42 \\

& $ 29 < m \le 30$ \nObjBin{28}	  & 0.89 & 0.68  & 0.71 & 0.32 \\

& $m > 30$ \nObjBin{5}	  & 0.60 & 0.40 & 0.20 & 0.00  \\

\bottomrule
\end{tabular}
\label{tab:spec_z}	
\end{table} 

In the fitting of the simulated NIRSpec spectra, which we describe in the next section, we assume a perfect knowledge of the spectroscopic redshift of the galaxies. Here we therefore discuss how well spectroscopic redshifts can be measured from deep NIRSpec/prism observations. To achieve this, we compute for each simulated spectrum the signal-to-noise ratio of the main UV and optical emission lines that NIRSpec can observe (see Fig.~\ref{fig:EL_visibility} for a list of lines). We then split the galaxies in the redshift and magnitude bins reported in Table~\ref{tab:spec_z}, and report in the same Table the fraction of objects in each bin for which we detect at least 1 or 2 emission lines with $\SN \ge 3$ and $\ge 5$. 

We consider that objects with at least 2 emission line detections with $\SN \ge 3$ or with at least 1 emission line detection with $\SN \ge 5$ will have secure spectroscopic redshifts, since in most cases \HST\ photometry will enable us to distinguish single line detections as being either \OIII\ or \Ha. Table~\ref{tab:spec_z} hence shows that out to redshift $z \sim 7$ the vast majority ($\gtrsim 90$ per cent) of the galaxies with $\magH \le 30$ will have secure spectroscopic redshifts, while this fraction drops to $\sim 70 $ per cent for fainter objects with $\magH > 30$. At $z \gtrsim 7$ the \Ha\ line is shifted outside the NIRSpec range, hence the fraction of galaxies with secure spectroscopic redshifts drops to $\sim 70 $ per cent for objects with $\magH \le 30$, while the number of galaxies in our catalogue with $\magH > 30$ and $z > 7$ is too low to provide a meaningful estimate of this fraction.
Overall, Table~\ref{tab:spec_z} demonstrates that only a few tens of objects in our catalogue may not have secure spectroscopic redshifts, i.e. $\sim 30$ per cent of $\magH > 30$ galaxies at $z \le 7$ and $\sim30$ per cent of $\magH > 28$ galaxies at $z > 7$.

\subsection{Fitting simulated \JWST/NIRSpec spectra}\label{sec:nirspec_fitting}

We use the \beagle\ tool to perform a full, pixel-by-pixel fitting of each simulated NIRSpec spectrum. We adopt a model similar to the one used to fit the \HST/XDF photometry (see Section~\ref{sec:fitting_XDF}), but fixing the redshift to the estimated value to mimic the way data will be analysed, i.e. with redshift being determined directly from the position of the spectral lines, or by combining single line detections with photometric redshifts (see Section~\ref{sec:spec_z} above). Unlike in the \HST\ broad-band fitting, we let the interstellar metallicity \Zism\ and fraction of $V$-band attenuation arising in the diffuse ISM, \mud, free to vary. We adopt the same independent priors as in Section~\ref{sec:fitting_XDF}, i.e. uniform for $\log (\M/\Msun)$, $\log (\tausfr / \txn{yr})$, $\log (Z / \Zsun)$, \logUs, $\xid$,  $\log (\Zism / \Zsun)$ and $\mu$, exponential for \tauV\ and Gaussian for $\log (\t / \txn{yr})$ and $\log (\sfrc / \MsunyrInv)$ (see Table~\ref{tab:XDF_priors}). As in Section~\ref{sec:fitting_XDF}, we consider a multi-variate Gaussian likelihood function, assuming independent errors in each pixel (see however Appendix~\ref{app:etc_limitations}). The likelihood function can therefore be described by equation~\eqref{eq:likelihood}, considering the summation over all spectral pixels, and assuming $\upsigma_i = \upsigmao$, i.e. removing any additional error term since, by construction, we have a perfect knowledge of the noise properties of the simulated spectra. We defer to a future work, based on the NIRSpec Instrument Performance Simulator, a more detailed analysis of other sources of random and systematic uncertainties in \JWST/NIRSpec data, related, for instance, to detector defects, flat-fielding and spectro-photometric calibration (see Appendix~\ref{app:etc_limitations}). 

We show in Fig.~\ref{fig:nirspec_spectra} two examples of the \beagle\ fitting of the simulated spectra. The red lines indicate the spectra obtained by computing, in each pixel, the median of the posterior distribution of predicted fluxes, while the red shaded regions show the 95 per cent central credible interval computed from the same distribution of fluxes. Fig.~\ref{fig:nirspec_spectra}(a) and (b) show how the high-\SN\ detections of several emission lines constrain the model predictions to high precision. As we will discuss in the next section, this in turn induces tight constrains on several physical quantities describing the conditions of photoionized gas in these galaxies.

\section{\JWST/NIRSpec constraints on galaxy properties}\label{sec:constraints}

\begin{figure*}
	\centering
	\resizebox{\hsize}{!}{\includegraphics{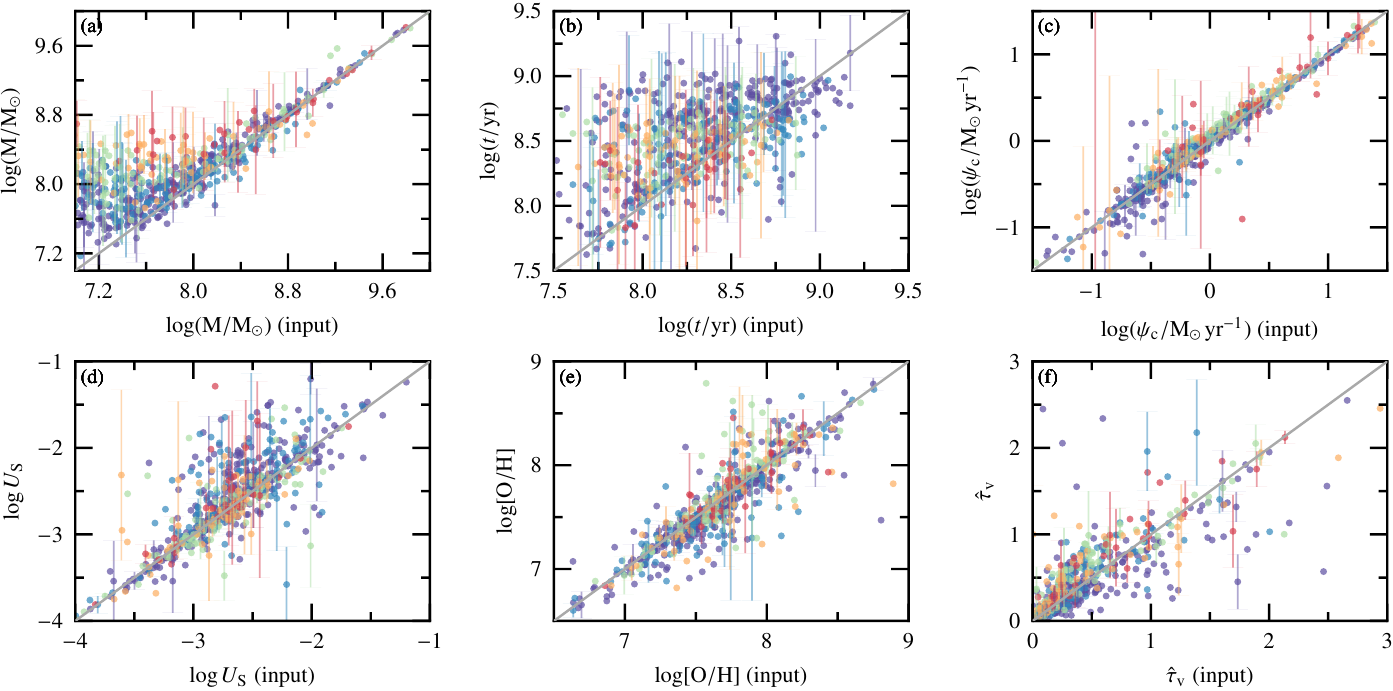}}
	\caption{Comparison between the input physical parameter (on the x-axis) and the value retrieved by fitting the full NIRSpec/prism simulated spectra with \beagle\ (on the y-axis) for the 6 model parameters stellar mass \logM, $\log (\t / \txn{yr})$, current star formation rate $\log (\sfrc / \MsunyrInv)$, ionization parameter \logUs, gas-phase metallicity \logOH\ and attenuation optical depth \tauV. The different colours correspond to the different dropout classes, and follow the colour coding of Fig.~\ref{fig:XDF_mag}. The retrieved parameter value corresponds to the posterior median of the marginal posterior distribution, while the error bars to the 68 per cent central credible region. To avoid overcrowding the plot, we only show errorbars for 20 randomly selected objects.}
	\label{fig:diff}
\end{figure*}


\begin{table*}
	\centering
\caption{Error thresholds on the retrieved model parameters corresponding to, 0.2, 0.5 and 0.9 of the inverse cumulative distribution function (quantile function) of the errors $\upsigma_\theta$, for the different dropout classes. The error $\upsigma_\theta$ is defined as half of the 68 per cent central credible interval ($1\upsigma$ for a Gaussian distribution). Each cell indicate the error threshold $s_\theta$ on the parameter $\theta$ such that the cumulative distribution $\mathcal{C}(\upsigma_\theta \le s_\theta)=0.2$, 0.5 and 0.9.}
	\begin{tabular}{c c c c c c c c c c c c c c c c}
\toprule


\multirow{2}{*}{Parameter}	     &  \multicolumn{5}{c}{ $s^{0.2}_\theta$}  &  \multicolumn{5}{c}{ $s^{0.5}_\theta$}  &  \multicolumn{5}{c}{ $s^{0.9}_\theta$}  \\     

\cmidrule(lr){2-6} \cmidrule(lr){7-11} \cmidrule(lr){12-16}

						  & $B$ & $V$ & $I$ & $Z$ & $Y$ &  			$B$ & $V$ & $I$ & $Z$ & $Y$ &	$B$ & $V$ & $I$ & $Z$ & $Y$ \\ 
  
\midrule

$\log (\M/\Msun)$		&            0.18 & 0.18 & 0.27 & 0.28 & 0.15 & 0.36 & 0.34 & 0.42 & 0.38 & 0.35 & 0.65 & 0.59 & 0.55 & 0.54 & 0.49  \\
                                                                                                                                           
$\log (\t / \txn{yr})$	    &    0.32 & 0.38 & 0.33 & 0.29 & 0.31 & 0.56 & 0.55 & 0.55 & 0.50 & 0.43 & 0.90 & 0.86 & 0.79 & 0.77 & 0.68  \\
                                                                                                                                           
$\log (\sfrc / \MsunyrInv)$		&  0.06 & 0.05 & 0.07 & 0.14 & 0.20 & 0.10 & 0.10 & 0.14 & 0.22 & 0.33 & 0.36 & 0.39 & 0.51 & 0.93 & 1.27  \\
                                                                                                                                           
\logUs		    		  &            0.05 & 0.05 & 0.05 & 0.07 & 0.05 & 0.15 & 0.18 & 0.12 & 0.15 & 0.16 & 0.50 & 0.59 & 0.47 & 0.74 & 0.73  \\
                                                                                                                                           
\logOH 		        		   &       0.04 & 0.04 & 0.05 & 0.07 & 0.06 & 0.09 & 0.10 & 0.09 & 0.13 & 0.19 & 0.23 & 0.37 & 0.33 & 0.49 & 0.58  \\
                                                                                                                                           
$\tauV$	    			    &          0.23 & 0.22 & 0.20 & 0.27 & 0.22 & 0.41 & 0.34 & 0.32 & 0.38 & 0.33 & 0.98 & 0.92 & 0.83 & 0.94 & 0.70  \\

\bottomrule
	\end{tabular}
\label{tab:cumul_errors}	
\end{table*}

We summarise in Fig.~\ref{fig:diff} and Table~\ref{tab:cumul_errors} the main results of our analysis of simulated `deep` NIRSpec observations of high redshift galaxies. We report the constraints on the 6 quantities stellar mass \logM, galaxy age $\log (\t / \txn{yr})$, current star formation rate $\log (\sfrc / \MsunyrInv)$, ionization parameter \logUs, gas-phase metallicity \logOH\ and $V$-band attenuation optical depth \tauV.  Fig.~\ref{fig:diff} shows in different panels the relation between retrieved and input value for the 6 quantities above. The large errorbars visible in Fig.~\ref{fig:diff}(a) in galaxies with $\log(\M/\Msun) \lesssim 8$ , along with the `flattening' of the relation between input and retrieved value, highlight the large uncertainty affecting the determination of the stellar masses, for which the posterior distribution is dominated by the prior. The stellar continuum emission in these galaxies, related to the stellar mass, is in fact too faint to be detected in our simulated spectra. Similarly, the large errorbars and spread of the points in Fig.~\ref{fig:diff}(b) show the difficulty of determining the age of the oldest stars from our simulated NIRSpec spectra. The reason is that the UV-to-optical spectra of galaxies with large specific star formation rates are dominated by the emission from young stars, which outshine the older ones. Fig.~\ref{fig:diff}(c), (d), (e) and (f) show that the star formation rate, ionization parameter, gas-phase metallicity and attenuation optical depth can be constrained across the entire range spanned by the input parameters. The constraints on these quantities come from emission lines ratios, and are therefore largely unaffected by our ability to detect the stellar continuum emission in faint, high-redshift galaxies.

We quantify in Table~\ref{tab:cumul_errors} the precision of the constraints on the 6 physical parameters shown in Fig.~\ref{fig:diff}. For this, we compute for each parameter the cumulative distribution of the errors $\upsigma_\theta$, and report in Table~\ref{tab:cumul_errors} the errors $s_\theta$ corresponding to 0.2, 0.5 and 0.9 of the inverse cumulative distribution $\mathcal{C}^{-1}$, i.e. $s^{0.2}_\theta$, $s^{0.5}_\theta$ and $s^{0.9}_\theta$. The values $s_\theta$ therefore indicate the error on the parameter such that 20, 50, or 90 per cent of the objects exhibit errors $\upsigma_\theta  \le s_\theta$. 
The central column of Table~\ref{tab:cumul_errors} indicates that for $\sim50$ per cent of the simulated galaxies the stellar masses (or mass-to-light ratios) can be constrained within a factor of $\sim2.5$ ($\sim 0.4$ dex), the age of onset of star formation within a factor of $\sim3.5$ ($\sim 0.55$ dex), the star formation rate, ionization parameter and gas-phase metallicity within a factor of $\sim 1.5$ ($\sim 0.2$ dex), albeit the precision worsens for galaxies at $z\gtrsim 7$ ($Z$ and $Y$ dropouts). The $V$-band dust attenuation optical depth can be constrained with a precision $\upsigma_{\tauV} \lesssim 0.4$.

The model parameters not reported in Fig.~\ref{fig:diff} and Table~\ref{tab:cumul_errors}, namely the timescale of star formation $\log (\tausfr / \txn{yr})$, dust-to-metal mass ratio $\xid$ and fraction of dust attenuation arising in the diffuse ISM \mud, are largely unconstrained by this analysis. As for the age of the oldest stars, constraining \tausfr\ and \mud\ would require detecting at high \SN\ the rest-frame optical-to-near infrared (stellar) continuum emission of the galaxies, which traces older stellar populations. Since the NIRSpec observations here considered mainly provide high-\SN\ detections of optical emission lines, a way to better constrain these parameters would be to combine NIRSpec spectra with \JWST/NIRCam and MIRI photometry. This type of analysis will be included in a future study. Constraining $\xid$, on the other hand, requires measuring, with high \SN, emission lines of refractory and non-refractory elements, such as oxygen (\OII, \OIII), and nitrogen and/or sulphur (\NII, \SII, e.g. see figure~3 of \citealt{Gutkin2016}). The emission lines of non-refractory elements, however, are too weak to be detected in the galaxies studied in this work. We note that although \tausfr, \mud\ and \xid\ are unconstrained by the simulated data, their inclusion in the fitting allows us to appropriately propagate into the constraints on the other physical parameters the uncertainty deriving from their poor knowledge, i.e.  \tausfr, \mud\ and \xid\ in this analysis play the role of `nuisance' parameters.

We now discuss the effect on the precision of the physical parameters retrieval of the \Ha\ line moving beyond the spectral coverage of NIRSpec at $z\gtrsim 7$ (see Fig.~\ref{fig:EL_visibility}). Table~\ref{tab:cumul_errors} shows that the error threshold $s^{0.5}_\theta$ for the parameters $\log \sfrc $ and \logOH\ remains constant or slowly increases from the $B$, to $V$, to $I$ dropouts, reflecting the presence of fainter objects, on average, when moving from lower to higher redshift sources (see Fig.~\ref{fig:XDF_mag}). The $Z$ and $Y$ dropouts, on the other hand, exhibit significantly larger $s^{0.5}_\theta$ values, likely because of the disappearance of the \Ha\ line from the NIRSpec data at those redshifts. These parameters are in fact mainly constrained by Balmer lines (\sfrc) or line ratios involving such lines [\logOH], and given that \Ha\ is intrinsically more luminous than \Hb, and that it suffers less attenuation, the constraining power of NIRSpec data suddenly drops as \Ha\ is shifted outside the NIRSpec range. The constraints on the ionization parameter \logUs, on the other hand, suffer less from the disappearance of \Ha, since the \OII\ line also provides constraints on this parameter.
The NIRSpec constraints on the stellar mass (or mass-to-light ratio) do not significantly evolve with redshift, since most galaxies in our catalogue exhibit low \SN\ in the continuum at all redshifts. A similar behaviour is observed for \tauV, likely because of the small absolute values of this parameter, and comparatively large retrieval errors, in our simulations. The age of onset of star formation $\log (\t / \txn{yr})$ is the parameter with the poorest constraints: as for \tausfr, constraining the galaxy age requires high \SN\ detections of the stellar continuum of evolved stellar populations, which is only possible for the brightest galaxies in the sample.

We note that the constrains on galaxy physical parameters discussed in this section do not account for model systematic uncertainties, since we have adopted the same physical model to create the mock spectra used as inputs for the NIRSpec simulations and to fit the simulated spectra. Physical models describing galaxy emission (e.g. stellar population and photoionization models) are built from the combination of several (uncertain) ingredients and assumptions, whose effect on the model predictions is difficult to quantify \citep[e.g. see][and references therein]{Conroy2013}. The constraints reported in Figure~\ref{fig:diff} and Table~\ref{tab:cumul_errors} hence must be considered as \emph{lower limits} on the \emph{precision} with which fundamental galaxy properties can be measured with deep \JWST/NIRSpec observations.

\section{Discussion}\label{sec:discussion}

The results presented in the previous section demonstrate how the unique combination of sensitivity and wavelength coverage of the NIRSpec/prism configuration will allow the precise characterisation of the physical properties of star-forming galaxies across a wide range of redshifts and luminosities. Specifically, the ability to observe multiple optical emission lines at high \SN\ should enable precise measurements of the gas-phase metallicity and star-formation rate, which, combined with stellar mass constraints from NIRCam, should significantly boost our ability to understand the mass and metal assembly in galaxies at redshift \range{z}{3}{10}. We note that a way to improve the precision of the physical parameter estimates of objects with low \SN\ emission line detections would be to combine them with galaxies with higher \SN\ measurements into a Bayesian hierarchical framework. This would enable population-wide constraints on the parameters \emph{and} to increase the precision of individual constraints. An ongoing effort to study such an approach is currently underway (Curtis-Lake et al., in prep). 

\begin{figure}
	\centering
	\resizebox{1.0\hsize}{!}{\includegraphics{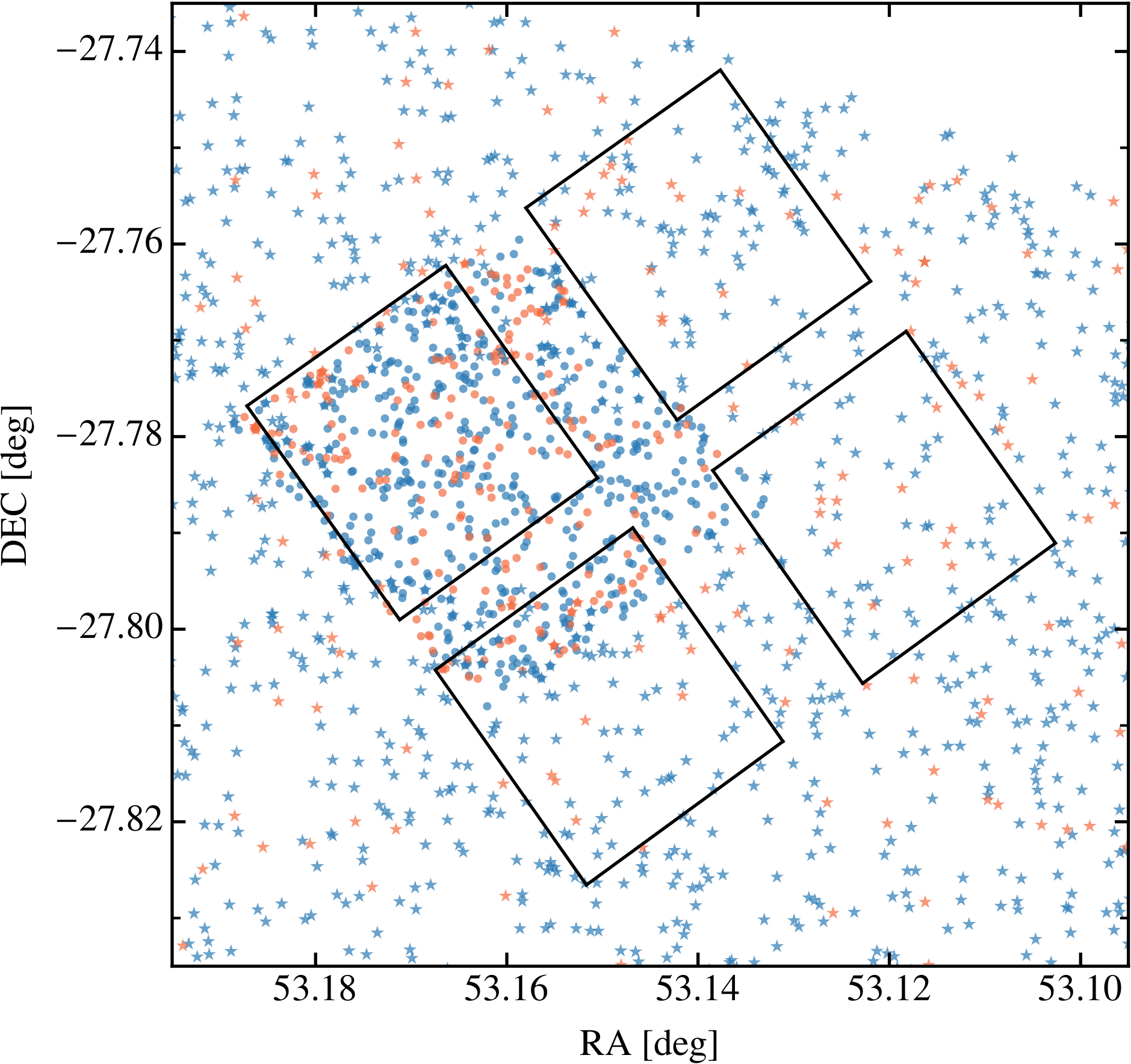}}
	\caption{The four quadrants of the NIRSpec micro-shutter array (MSA) overplotted on galaxies at redshift $4 \lesssim z \lesssim 6$ (blue symbols) and $z\gtrsim 6$ (orange symbols) selected by \citet{Bouwens2015} in the HUDF/XDF (circles) and CANDELS/GOODS-South `deep' (stars) regions.}
	\label{fig:NIRSpec_pointing}
\end{figure}

The analysis presented so far illustrates the statistical constraints on several galaxy properties that can be obtained from 100 ks NIRSpec/prism observations of high-redshift sources, but it does not provide any insight on the actual number of sources that will be observed with a NIRSpec pointing. Here, we therefore discuss the number of \HST-selected galaxies expected to be observed in a single NIRSpec/prism pointing centred on the HUDF. Fig.~\ref{fig:NIRSpec_pointing} shows the HUDF/GOODS-South area along with the four NIRSpec MSA quadrants, where we centred one quadrant on the 4.6 \sqarcmin\ area of the HUDF with deep WFC3 observations. Fig.~\ref{fig:NIRSpec_pointing} shows how HUDF sources will only be observable from one MSA quadrant, while for the remaining three quadrants galaxies will have to be selected from shallower data in GOODS-South.\footnote{The HUDF/XDF area is larger than the on-sky area of a single MSA quadrant ($1.58\times1.40$ arcmin, excluding vignetted shutters), but the separation between quadrants ($0.38$ arcmin along one axis, $0.62$ arcmin along the other) and unknown telescope roll-angle will in practice limit the possibility of selecting HUDF/XDF targets in more than one quadrant.}
We report in Table~\ref{tab:pointings} the expected number \npt\ of galaxies at \range{z}{4}{5} and $z\gtrsim6$ that can be observed with a single NIRSpec pointing in the HUDF/GOODS-South area, along with the on-sky density of galaxies in the HUDF/XDF and GOODS-South regions. Because the faintest targets in the HUDF/XDF may not be detected at a significant level in our deep NIRSpec/prism observations, we also report the expected number of targets observed in the HUDF/XDF with $\magH < 29.5$ and $<29$. From the on-sky density of the targets, and assuming Poisson-distributed sources, a 3 shutters slitlet configuration, and allowing targets to occupy any position within the open area of a shutter (Ferruit at al., in prep), we can compute the number of expected sources observed in a single NIRSpec pointing. At $z\gtrsim6$, we consider $\sim 200$ potential targets in the HUDF/XDF and $\sim 300$ in CANDELS/GOODS-South, which implies $\sim 30$ non-overlapping galaxy spectra observed, on average, in a single pointing. At \range{z}{4}{6}, the larger number of targets, $\sim 500$ in the XDF and $\sim 2000$ in CANDELS/GOODS-South, translates into $\sim 60$ spectra observed in a single pointing. This will leave $\sim 100$ unused slitlets which will be occupied by galaxies at lower redshift. 
Table~\ref{tab:pointings} also indicates that restricting the $z\gtrsim6$ targets in the HUDF/XDF to $\sim 80$ galaxies with $\magH \le 29$ (which should exhibit $\SNHb>5$, see Table~\ref{tab:Hb_SN}), reduces by $1/3$ (from $\sim 30$ to $\sim 20$) the number of observed galaxies in a pointing. Given the paucity of $z\gtrsim6$ targets, it is therefore advisable to prioritise the brighter targets without excluding the fainter ones, as we have previously shown that a significant fraction of $\magH > 29$ galaxies should show $\SN>3$ detections of \Hb\ (see Table~\ref{tab:Hb_SN}).
We note that the main factor limiting the NIRSpec multiplexing of $z\gtrsim6$ galaxies is their on-sky density outside the HUDF/XDF region, as this density is only $\sim5\,\invsqarcmin$ in the CANDELS/GOODS-South `deep' area, where data reach a $5 \, \upsigma$ depth of $\magH \sim 27.5$.

\begin{table}
\begin{threeparttable}
\centering
\caption{Number of galaxies at \range{z}{4}{6} and $z\gtrsim 6$ expected to be observed with a single NIRSpec/prism pointing centred on the HUDF.}
	\begin{tabular}{C{0.195\columnwidth-2\tabcolsep} C{0.14\columnwidth-2\tabcolsep} C{0.12\columnwidth-2\tabcolsep} C{0.135\columnwidth-2\tabcolsep} C{0.115\columnwidth-2\tabcolsep} C{0.12\columnwidth-2\tabcolsep} C{0.125\columnwidth-2\tabcolsep} C{0.115\columnwidth-2\tabcolsep}}
\toprule


\multirow{2}{=}{\centering Catalog}	     & \multirow{2}{=}{\centering \magH}	     & \multicolumn{3}{c}{\range{z}{4}{6}\tnote{a}}  &  \multicolumn{3}{c}{$z\gtrsim6$\tnote{b}}  \\     

	\cmidrule(lr){3-5} \cmidrule(lr){6-8} 

				  & 	& \Nobj	& \Sobj\tnote{c}	& \npt\tnote{d}	& \Nobj	& \Sobj \tnote{c}	& \npt\tnote{d} 	\\
  
\midrule

\multirow{3}{=}{\centering XDF\tnote{e}}  & $< 29$	& 304	& $\sim65$	& $\sim 18$	& 80	 	& $\sim 17$	&	$\sim 11$ \\ 

				  & $< 29.5$	& 413	& $\sim 88$	& $\sim 17$	& 132	& $\sim 28$	&	$\sim 16$ \\

				  & all		& 513	& $\sim 110$	& $\sim 15$	& 202	& $\sim 43$	& 	$\sim 21$ \\
				 
\addlinespace[6pt]
				  
GOODS-S\tnote{f} & all	&  2061	& $\sim 32$	& $\sim 46$	& 303	& $\sim 4.7$	& $\sim 10$	\\

\bottomrule
\end{tabular}
\begin{tablenotes}
\item [a] For XDF, we include the $B$ and $V$ dropouts. For CANDELS, we consider galaxies at $z \sim 4$ and $z \sim 5$ from table~4 of \citet{Bouwens2015}. 
\item [b] For XDF, we include the $I$, $Z$ and $Y$ dropouts. For CANDELS, we consider galaxies at $z \gtrsim 6$ from the same table as in \tnote{a}.
\item [c] On-sky target density of the objects in units of \invsqarcmin.
\item [d] Average number of non-overlapping spectra observed with the NIRSpec/prism in a single pointing.
\item [e] The number of objects $\txn{N}_\txn{obj}$ refers to the entire XDF area of $\sim4.7 \, \sqarcmin$, while to compute $\txn{n}_\txn{p}$ we consider 1 MSA quadrant.
\item [f] We consider the total number of objects $\txn{N}_\txn{obj}$ in CANDELS/GOODS-South `deep', corresponding to an area of $\sim 64.5 \, \sqarcmin$
\end{tablenotes}
\label{tab:pointings}	
\end{threeparttable}
\end{table}

Our previous analysis has shown that `deep' NIRSpec/prism observations will be able to constrains fundamental properties of $z\gtrsim6$ galaxies, such as their gas-phase metallicities and star formation rates (Sec~\ref{sec:constraints}). The above discussion indicates, however, that obtaining a statistically significant sample of objects with precise measurements of these properties will require several `deep', 100~ks, pointings centred on the deepest \HST/WFC3 fields, later exploiting targets selected from \JWST/NIRCam imaging.
 
\section{Summary and conclusions}\label{sec:conclusions}

In this work, we have presented statistical constraints on different galaxy physical properties obtained from the analysis of simulated NIRSpec/prism observations of high-redshift sources. We have introduced an original approach to build a semi-empirical `mock' catalogue of $z\gtrsim3$ galaxy spectra, based on observed photometric SEDs of a sample of $\sim700$ galaxies identified as $B$-, $V$-, $I$-, $Z$- and $Y$-band dropouts in the \Hubble\ Ultra Deep Field. We used the state-of-the-art \beagle\ tool to match the observed UV-to-near infrared broad-band SED of each galaxy with a diversity of model spectra statistically consistent with the \HST\ photometry. We showed how additionally requiring that the models follow a (redshift-dependent) relation between stellar mass and star formation rate reduces degeneracies among model parameters, allowing us to relate, in a statistical way, model spectra with \HST-selected galaxies. We validated the above approach by comparing the predicted \Spitzer/IRAC band-1 (3.6 \micron, \IRACone) and band-2 (4.5 \micron, \IRACtwo) fluxes of the resulting mock catalogue of galaxy spectra with observed stacked photometry of galaxies at $z\sim4$ and $z\sim5$, and with the evolution of the IRAC colour \IRACcolour\ at redshift \range{z}{2}{7}. We also showed that our model predictions cover the entire observed range of \CIII\ equivalent widths at \range{z}{1}{8}.

We have exploited the mock galaxy catalogue described above to simulate \JWST/NIRSpec spectra using an `Exposure Time Calculator'-like (ETC-like) approach. Specifically, we considered NIRSpec observations performed with the low-resolution prism, which covers the wavelength range \range{\lambda}{0.6}{5.3} \micron, assuming a total integration time of $\sim100$~ks. Such an observational strategy is tailored to the detection of faint, distant galaxies out to $z\sim10$. Our simulations show that this strategy enables the detection of several optical emission lines (e.g. \Hb, \Ha, \OIII) for the majority of \HST-selected targets out to $z\gtrsim8$. In particular, we expect a $\SN>3$ detection of the \Hb\ line in most $z\gtrsim3$ galaxies brighter than $\magH \sim 30$, and a significant fraction of detections of \Hb\ in $\magH > 30$ objects. Our calculations also imply $\SN \gtrsim 20$ detections of the \Ha\ and \OIII\ lines for most sources in the HUDF.

We also used the \beagle\ tool to analyse the simulated NIRSpec spectra. We could show in this way that the simultaneous detection of several optical emission lines enables tight constraints on the star formation rate, dust attenuation and ionised-gas properties in $z\gtrsim3$ galaxies. Our analysis indicates that such observations will allow measurements of star formation rates, ionization parameters and gas-phase metallicities within a factor of $\sim 1.5$, mass-to-light ratios within a factor of $\sim2$, galaxy ages within a factor of $\sim3$ and $V$-band attenuation optical depths with a precision of $\sim0.3$. We also showed how the shift of the \Ha\ line outside the NIRSpec wavelength range at $z\gtrsim7$ lowers the precision of the constraints on most physical parameters, especially star formation rate, gas-phase metallicity and ionization parameter. 

Finally, we have discussed our findings in the context of a NIRSpec/prism pointing centred on the HUDF. We showed that the small area covered by deep \HST/WFC3 observations (4.7 \sqarcmin) and the low on-sky density of high-redshift targets in the CANDELS/GOODS-South region limit the expected number of $z\gtrsim 6$ galaxies that can be observed in a single NIRSpec pointing. We have shown that, on average, a single NIRSpec pointing, covering an effective (non-contiguous) area of $\sim9.8$ \sqarcmin, will enable the observation of $\sim20$ ($\sim15$) galaxies at $z\gtrsim6$ (\range{z}{4}{6}) selected from the HUDF/XDF region, and $\sim10$ ($\sim45$) galaxies at $z\gtrsim6$ (\range{z}{4}{6}) from the shallower CANDELS/GOODS-South one. This therefore indicates that multiple `deep' NIRSpec/prism pointings targeting known high-redshift galaxies will be necessary to assemble a statistically significant sample of $z\gtrsim4$ galaxy spectra with high-\SN\ emission-line detections.

In the future, we plan to extend our simulations to account for different observational strategies, in particular considering the combination of low-, medium- and high-spectral resolution configurations used in the NIRCam and NIRSpec Guaranteed Time Observer teams for their galaxy assembly survey \citep{Rieke2017}.
The simulations presented in this work will be made public on the Github repository \url{https://github.com/jacopo-chevallard/Simulating-deep-NIRSpec-observations-in-the-Hubble-Ultra-Deep-Field}.

\section*{Acknowledgements}

The authors thank the anonymous referee for a careful reading of the manuscript which significantly improved the final version of the paper. JC, ECL and S. Charlot acknowledge support from the European Research Council (ERC) via an Advanced Grant under grant agreement no. 321323-NEOGAL. S. Carniani acknowledges financial support from the UK Science and Technology Facilities Council (STFC). SA is funded by the Spanish Ministerio de Econom\'ia y Competitividad (MINECO) under grant ESP2015-68964-P. RS acknowledges support from a NWO Rubicon program with project number 680-50-1518. RM and RA acknowledge support from the ERC Advanced Grant no. 695671-QUENCH.

\bibliographystyle{mnras}

\bibliography{NIRSpec_simulations} 

\appendix

\section{Generating simulated NIRSpec spectra}\label{app:etc_simul}

In this Appendix, we describe the different steps of the approach that we used to simulate NIRSpec spectra.

\subsection{Exposures and observing strategy}

In this work, we use the standard exposure parameters and observational strategy baselined to monitor the NIRSpec performances during the instrument development. For the individual exposures, we consider the MULTIACCUM readout pattern $22\times4$ listed in Table~2 of \citet{Rauscher2007}, which yields an effective integration time of $t_\txn{eff}=934.38~\txn{s}$ per-exposure. Each exposure lasts 955.86~s, including the time needed to reset the array at the beginning of the exposure. The simulations are performed for a total of $n_\txn{exp}=108$ exposures, corresponding to a cumulated effective integration time of 100923.84~s, slightly in excess of 100~ks.

The building block of a standard NIRSpec MOS observation of faint objects is a minimum set of 3 nodding-pattern exposures in which the object is observed consecutively in three neighbouring micro-shutters forming a $1\times3$ slitlet. For each of the 3 object positions, two measurements of the telescope and zodiacal light backgrounds are available and can be used for an efficient and accurate pixel-level background subtraction. This implies considering the average of $n_\txn{b}=2$ observations, over the same pixels used for the object spectrum, to estimate the contribution of the background subtraction to the total variance spectrum.

The effective integration time is further reduced by a factor $\gamma_\txn{bad pixels} = 0.807$ to account for bad pixels. The factor $\gamma_\txn{bad pixels}$ is estimated by computing the probabilities of loosing one, two or three measurements in a 3-slitlet nodding pattern, assuming a fraction of bad pixels of 3.5 per cent. Accounting for bad pixels hence yields a final integration time of $t_\txn{final} = t_\txn{eff} \times \gamma_\txn{bad pixels} = 754.04~\txn{s}$ per-exposure. We note that the fraction of bad pixels here considered is conservative, as it corresponds to end-of-life values. The official \JWST\ exposure time calculator (`Pandeia' ETC; \citealt{Pontoppidan2016}) does not account for this effect, and this can introduce differences among our simulations and those obtained with the \JWST\ ETC, at fixed observational setting. We do not model the impact of cosmic rays, as this is expected to be a minor effect for the short individual exposures that we have adopted.

\subsection{Extraction of the spectrum}

As we simulate integrated, 1-dimensional observed spectra, we must also consider a spectrum extraction strategy. We assume a simple extraction strategy by summing over $n_\txn{spa}=4$ detector pixels along the spatial direction, ignoring any light that may be lost in the process. More accurate simulations, based on the NIRSpec Instrument Performance Simulator, show that this assumption is correct within a few percent for a point source. No summation was performed along the spectral direction ($n_\txn{spe}=1$).

\subsection{Electron rates}\label{app:e_rates}

In order to estimate the noise variance associated to each element of the simulated spectrum, we need to compute the number of electrons accumulated during an exposure, for both the object and the background (zodiacal light + parasitic light from the telescope) spectra.
For the object, we start from a high-resolution, spatially integrated, redshifted spectrum $\txn{F}^{o}(\lambda)$ in units of $\txn{W}\,\txn{m}^{-2}\,\txn{m}^{-1}$ (spectral irradiance) generated with the \beagle\ tool as detailed in Section~\ref{sec:semi_empirical}. By linearly rebinning the spectrum, we construct an input spectrum $R^{o}(k)$ in units of $\txn{W}\txn{m}^{-2}$ on the pixel grid ($n_k = 365$ typically) representative of the NIRSpec PRISM/CLEAR configuration. In this process, we take into account the highly variable spectral resolution of the NIRSpec/prism ($\R \sim 30$ to $\sim 300$), and consider the wavelength range $0.6-5.0~\micron$. Note that there is no one-to-one correspondence between the pixels of the rebinned input spectrum and the actual detector pixels, as the former contain the spatially integrated irradiance of the object, i.e. after summation over 4 detector pixels along the spatial direction.

Note that in this approach we do not convolve the input spectrum by the (wavelength-dependent) line-spread function of the instrument. As the emission lines in the input spectrum are spectrally unresolved at the NIRSpec/prism spectral resolution, this causes them to be unrealistically concentrated into single pixels in our rebinned spectra. This yields, on average, a $\sim35$ per cent larger emission line \SN\ in our simulated spectra with respect to a computation accounting for the spectral response. As highlighted in section~\ref{app:etc_limitations}, this overestimation of the peak signal-to-noise ratio of the emission lines is largely compensated by the conservative values adopted in other steps of our computations (see Table~\ref{tab:effects}).

The number of electrons $\txn{N}^{o}(k)$ accumulated during an individual exposure and associated to the object is computed using the relation
\begin{equation}
\txn{N}^{o}(k) = \frac{R^{o}(k) \times t_\txn{final} \times \txn{A}_\txn{telescope} \times \txn{PCE}(\lambda_{k}) \times \eta^{o}(\lambda_{k})}{\txn{E}(\lambda_k)} \, ,
\end{equation}
where $\lambda_{k}$ is the central wavelength of the $k^{th}$ spectrum pixel (in m), $\txn{A}_\txn{telescope}$ collecting area of the telescope ($25~m^2$), $\txn{PCE}(\lambda_k)$ the photon conversion efficiency of the telescope + instrument (in units of electrons-per-collected-photon), $\eta^{o}(\lambda_k)$ the slit losses factor (unitless, see Appendix~\ref{app:slit_losses}) and $\txn{E}(\lambda_k)=hc/\lambda_k$ the energy-per-photon (in $J\,\txn{photon}^{-1}$). The photon conversion efficiency used in these simulations is consistent with the one used in the official \JWST\ ETC version 1.1.1, available in October 2017.

The number of accumulated electrons $\txn{N}^{b}(k)$ for the background (zodiacal light + parasitic telescope light) is computed using the relation
\begin{equation}
\txn{N}^{b}(k) = \frac{\txn{F}^{b}(\lambda_k) \times \delta\lambda(k) \times t_\txn{final} \times \txn{A}_\txn{telescope}  \times \txn{PCE}(\lambda_{k}) \times \Omega_\txn{sky}}{\txn{E}(\lambda_k)} \, ,
\end{equation}
Where $\txn{F}^{b}(\lambda_k)$ is the spectral radiance (surface brightness) of the background in units of $\txn{W}\,\txn{m}^{-2}\,\txn{m}^{-1}\,\txn{arcsec}^{-2}$, $\delta\lambda(k)$ the size of the current spectrum pixel in units of wavelength (in m) and $\Omega_{sky}$ the projected on-sky area of a background element (in $\txn{arcsec}^{2}$). The area $\Omega_{sky}$ is computed considering, along the spectral direction, the width of a micro-shutter ($0.20~\txn{arcsec}$), and 4 detector pixels ($0.1~\txn{arcsec}$ each) along the spatial direction. For the surface brightness of the zodiacal light, we use a simple model which combines a power law and a $256~\txn{K}$ black-body spectrum. We then scale this zodiacal light model to match the typical spectral radiance at the north ecliptic pole, then multiply it by a factor 1.2, as customary done to evaluate the NIRSpec performances. For the contribution of the telescope straylight and background, we use constant values of $0.091~\txn{MJy}\,\txn{sr}^{-1}$ below $2~\micron$ and $0.07~\txn{Mjy}\,\txn{sr}^{-1}$ above $3~\micron$, with a linear interpolation in between. These values correspond to the requirements placed on the observatory at 2 and 3 \micron, and are conservative levels when compared with the expected background in Hubble Deep Field South region \citep[e.g.][]{Wei2006}.

\subsection{Noise model}

We use a simplified noise noise model in which the total spectrum variance is computed considering two noise sources:
\begin{itemize}
\item shot noise of the object, of the background and of the detector dark current (assuming a conservative dark current rate of $d = 0.01~\txn{electron}\,\txn{s}^{-1}\,\txn{pixel}^{-1}$);
\item detector noise ($\upsigma_{r}$), assuming a conservative value of $7~\txn{electrons}\,\txn{pixel}^{-1}$.
\end{itemize}
The total variance of an individual exposure $\txn{V}(k)$ is then computed using the relation
\begin{equation}\label{eq:tot_variance}
\txn{V}(k) = \txn{N}^{o}(k) + \left( 1 + \frac{1}{\txn{n}_b} \right) \times \left[\txn{N}^{b}(k) + \txn{n}_\txn{spa}~\txn{n}_\txn{spe} \times \left(d~t_\txn{eff} + \upsigma_{r}^2 \right) \right] \, ,
\end{equation}
where $\txn{N}^{o}(k)$ is the number of accumulated `object' electrons per-exposure, $\txn{n}_b$ the number of background spectra available per-exposure, $\txn{N}^{b}(k)$ the number of accumulated `background' electrons, $\txn{n}_\txn{spa}$ and $n_\txn{spe}$ the number of detector pixels along the spatial and spectral directions used in the spectrum extraction, $d$ the dark current rate, $t_\txn{eff}$ the effective integration time and $\upsigma_{r}$ the detector noise rate.

\subsection{Generating the simulated spectrum}

After computing the variance, we can obtain the simulated, noisy spectrum $\txn{S}^{o}(k)$ (in $\txn{W}\,\txn{m}^{-2}\,\txn{m}^{-1}$) assuming a perfect background subtraction (i.e. no residuals other than the noise):
\begin{equation}
\txn{S}^{o}(k) = \frac{R^{o}(k)}{\delta\lambda(k)} + \sqrt{\frac{\alpha^{2}(k) \times \txn{V}(k)}{\txn{n}_\txn{exp}}} \times \mathcal{N}_k(0,1) \, ,
\end{equation}
where $R^{o}(k)$ is the rebinned (noiseless) spectrum, $\delta\lambda(k)$ the size of the spectrum pixel, $\alpha(k)$ the conversion factor from electrons to $\txn{W}\,\txn{m}^{-2}\,\txn{m}^{-1}$, $\txn{V}(k)$ the total variance, $\txn{n}_\txn{exp}$ the number of exposures and $\mathcal{N}_k(0,1)$ a random Gaussian variate drawn from a distribution with zero mean and unit standard deviation.

Note that we have also considered additional noise terms not shown in equation~\eqref{eq:tot_variance} to mimic the effects of a non-ideal detector dark-current subtraction (10 per cent accuracy, $1\upsigma$ per pixel, no correlation from one pixel to the next) and of a pixel-to-pixel flat-field correction (20 per cent accuracy, $1\upsigma$ per pixel, no correlation from one pixel to the next). These two contributions have, however, very little impact on the simulated spectra for the dark current and signal-to-noise ratio levels considered in this paper.

\subsection{Caveats and limitations}\label{app:etc_limitations}

The present work is based on simulations of NIRSpec/prism observations computed with a simple ETC-like approach. The flexibility and versatility of this approach allows us to easily explore different observational strategies (e.g. combination of spectral configurations, depth of observations) but it is based on some simplifying assumptions about the NIRSpec instrument. In Table~\ref{tab:effects}, we list various effects that can affect the signal-to-noise ratio and calibration of NIRSpec spectra, and indicate whether these effects are accounted for in our simulations and in the official \JWST\ ETC \citep{Pontoppidan2016}. Table~\ref{tab:effects} shows that, depending on the effect considered, our simulations can be optimistic or pessimistic. Accounting for these effects, however, would require significantly more complex and computationally-intensive simulations, for instance based on the NIRSpec Instrument Performance Simulator, and more realistic data reduction strategies.



\begin{table*}
\begin{threeparttable}
	\centering
	\caption{Summary of the main instrumental effects affecting simulated NIRSpec spectra computed with the approach adopted in this work, and with the official \JWST/NIRSpec ETC.}
	\begin{tabular}{C{0.18\textwidth-2\tabcolsep} C{0.26\textwidth-2\tabcolsep} C{0.26\textwidth-2\tabcolsep}  L{0.30\textwidth-2\tabcolsep} }
\toprule


\multicolumn{1}{c}{Effect}	    &  \multicolumn{1}{c}{This paper}  &  \multicolumn{1}{c}{\JWST\ ETC} &  \multicolumn{1}{c}{Impact on our simulations}  \\     

\midrule

MSA slit-losses		& Averaged & Yes &  See Appendix~\ref{app:slit_losses}\\

Line spread function	& No & Yes &  Significant impact on the peak signal-to-noise ratio of emission lines (optimistic, average effect of 35 per cent, linear impact)\\

Background	& Included in the noise model; conservative level; perfect subtraction &  Included in the noise model; adjusts level to on-sky position and observation time; perfect subtraction & Limited impact since background shot noise is only dominant in a small wavelength range around $1.2~\micron$ \\

Detector noise model		& Conservative total noise value (7 instead of nominal 6 electrons) & According to \citet{Rauscher2007} formalism & Significant impact on sensitivity limits (pessimistic, linear impact at long wavelengths)\\

Noise correlation & No & No (capability not activated for NIRSpec) & Impact not evaluated. Note that a readout mode ($\txn{IRS}^2$) optimised to reduce correlated noise is available for NIRSpec. \tnote{a}\\

Detector defects		& Yes (conservative end-of-life values) & No &  Significant impact on our simulations compared to ETC ones, as we consider that almost 20 per cent of the exposure time is lost due to bad pixels\\

Cosmic rays		& No & Yes &  Only minor impact as we are considering relatively short exposures\\

Dark current	& Included in the noise model; conservative rate of 0.01 $\txn{e}\,\txn{s}^{-1}$; non-perfect subtraction  & Included in the noise model; measured (lower) rates; perfect subtraction &  Limited impact as even conservative values of dark-current are very low\\

Flat fielding	& Non-ideal & Non-ideal &  Negligible impact as we assume that dithering will be used to mitigate this problem\\

Spectro-photometric calibration		& Perfect & Perfect &  Impact on retrieval of galaxy properties not evaluated \\

Wavelength calibration		& Perfect & Perfect & No impact \\

\bottomrule
	\end{tabular}
    \begin{tablenotes}
\item [a] \url{https://jwst-docs.stsci.edu/display/JTI/NIRSpec+IRS2+Detector+Readout+Mode}
\end{tablenotes}
\label{tab:effects}	
\end{threeparttable}
\end{table*}

\section{Slit-losses from the NIRSpec Micro Shutter Array (MSA)}\label{app:slit_losses}

A standard NIRSpec MSA mask will consist of a series of 1$\times$3 shutter `microslits', with the object of interest placed at the center of each shutter as part of a standard nodding scheme.  However, the extended size of an object compared to the small ($0.20'' \times 0.46''$) shutters, the bars between shutters, difficulties in perfectly centring every object within the shutters, and the large PSF at long wavelengths ensures that not all of the incident flux from a galaxy will make it onto the detector.  In order to quantify the effect of these sources of flux losses, we create simulations based on models of the instrument and models of galaxy light distributions.

In order to average over all source position angles and ellipticities, throughout we assume a round source with a S\'ersic light profile.  For a grid of S\'ersic indices and half-light radii, an object is positioned in multiple locations inside the 1$\times$3 microslit and the losses from both geometry and diffraction are calculated as a function of wavelength within the full aperture.  These calculations include all of the changes to the PSF as the light propagates to the MSA plane, including the Optical Telescope Element (OTE) and the NIRSpec foreoptics.  

Since the specifics of the mask layout and the exact positioning of objects within slits is not known, we calculate the median throughput as a function of wavelength for an object with a given size and S\'ersic index, adopting the APT centring tolerances, e.g. accepting that the object before nodding is within the full pitch of the central shutter (including the bars between shutters), accepting that the object is in the open area, etc.  We include the effect of nodding within the 1$\times$3 microslit aperture in these calculations, where more light is lost at the upper and lower nodding positions.  Since the quantity of interest is the relative throughput with respect to a centred point source, we correct our estimates according to the wavelength-dependent transmission of a centred point source as calculated using the same framework.


\bsp	
\label{lastpage}
\end{document}